\documentclass[twocolumn,amsmath,amssymb,floatfix,superscriptaddress]{revtex4-2}
\usepackage{dcolumn}
\usepackage{graphicx,xcolor}
\usepackage{dsfont}
\usepackage{ulem}

\newcommand{\ifn}{Istituto di Fotonica e Nanotecnologie - Consiglio Nazionale delle Ricerche (IFN-CNR), p.za Leonardo da Vinci 32, 20133 Milano, Italy}
\newcommand{\cnn}{Centre for Nanosciences and Nanotechnologies, CNRS, Universit\'e Paris-Saclay, UMR 9001, 10 Boulevard Thomas Gobert, 91120, Palaiseau, France}
\newcommand{\spz}{Dipartimento di Fisica, Sapienza Universit\`a di Roma, P.le Aldo Moro 5, 00185, Rome, Italy}
\newcommand{\univie}{University of Vienna, Faculty of Physics, Boltzmanngasse 5, 1090 Vienna, Austria}
\newcommand{\qdl}{Quandela SAS, 7 Rue L\'eonard de Vinci, 91300 Massy, France}
\newcommand{\pav}{Dipartimento di Fisica, Universit\`a di Pavia, Via Bassi 6, 27100, Pavia, Italy}
\newcommand{\usc}{Physics Department, Federal University of São Carlos, São Carlos 13565-905, SP, Brazil}

\usepackage{listings}
\usepackage{color}
\usepackage{graphicx, caption}
\usepackage{subcaption}
\usepackage{braket}
\usepackage{physics}
\usepackage[colorlinks=true]{hyperref}
\usepackage{array}
\newcolumntype{P}[1]{>{\centering\arraybackslash}m{#1}}

\usepackage{bbold}
\definecolor{mygreen}{RGB}{38, 254, 136}

\begin{document}
	
	\title{High-fidelity generation of four-photon GHZ states on-chip}
	
	\author{Mathias Pont}
    \thanks{These authors contributed equally to this work.}
	\affiliation{\cnn}	

	\author{Giacomo Corrielli}
    \thanks{These authors contributed equally to this work.}
	\affiliation{\ifn}

	\author{Andreas Fyrillas}
	\affiliation{\qdl}
	
	\author{Iris Agresti}
	\affiliation{\spz}
	\affiliation{\univie}
	
	\author{Gonzalo Carvacho}
	\affiliation{\spz}
	
	\author{Nicolas Maring}
	\affiliation{\qdl}
	
	\author{Pierre-Emmanuel Emeriau}
	\affiliation{\qdl}
	
	\author{Francesco Ceccarelli}
	\affiliation{\ifn}
	
	\author{Ricardo Albiero}
	\affiliation{\ifn}
	
	\author{Paulo Henrique Dias Ferreira}
    \affiliation{\usc}
	\affiliation{\ifn}
	
	\author{Niccolo Somaschi}
	\affiliation{\qdl}
	
	\author{Jean Senellart}
	\affiliation{\qdl}
	
	\author{Isabelle Sagnes}
	\affiliation{\cnn}
	
	\author{Martina Morassi}
	\affiliation{\cnn}
	
	\author{Aristide Lema\^itre}
	\affiliation{\cnn}
	
	\author{Pascale Senellart}
	\affiliation{\cnn}
	
	\author{Fabio Sciarrino}
	\affiliation{\spz}
	
	\author{Marco Liscidini}
	\affiliation{\pav}
	
	\author{Nadia Belabas}
    \thanks{These authors both closely supervised the work.}
	\affiliation{\cnn}	
	
	\author{Roberto Osellame}
    \thanks{These authors both closely supervised the work.}
	\affiliation{\ifn}
	
	\date{\today}
	
	\begin{abstract}
		Mutually entangled multi-photon states are at the heart of all-optical quantum technologies. While impressive progresses have been reported in the generation of such quantum light states using free space apparatus, high-fidelity high-rate on-chip  entanglement generation is crucial for future scalability. In this work, we use a bright quantum-dot based single-photon source to demonstrate the high fidelity generation of 4-photon Greenberg-Horne-Zeilinger (GHZ) states with a low-loss reconfigurable glass photonic circuit. We reconstruct the density matrix of the generated states using full quantum-state tomography reaching an experimental fidelity to the target $\ket{\text{GHZ}_4}$ of $\mathcal{F}_{\text{GHZ}_4}=(86.0\pm0.4)\,\%$, and a purity of $\mathcal{P}_{\text{GHZ}_4}=(76.3\pm0.6)\,\%$. The entanglement of the generated states is certified with a semi device-independent approach through the violation of a Bell-like inequality by more than 39 standard deviations. Finally, we carry out a four-partite quantum secret sharing protocol on-chip where a regulator shares with three interlocutors a sifted key with up to 1978 bits, achieving a qubit-error rate of 10.87\%. These results establish that the quantum-dot technology combined with glass photonic circuitry for entanglement generation on chip offers a viable path for intermediate scale quantum computation and communication.
	\end{abstract}
	
	\maketitle
	
	Entangled multi-partite states have a pivotal role in quantum technologies based on multiple platforms, ranging from trapped-ions~\cite{roos2004} to superconducting qubits~\cite{dicarlo2010}. In photonics, over the past two decades, major advances in the generation of multi-photon entanglement have been achieved by exploiting spontaneous parametric down-conversion (SPDC) and free-space apparatuses~\cite{pan2001, walther2005, kiesel2005, zhong2018, wang2018}. Very recently, one dimensional linear cluster states were generated on demand through atom-photon entanglement~\cite{thomasP2022} or spin-photon entanglement ~\cite{cogan2021, coste2022} at high rate~\cite{thomasP2022, coste2022} and high indistinguishability ~\cite{cogan2021, coste2022}. While the complexity of the generated states in some of these works~\cite{zhong2018, wang2018, thomasP2022} is still unmatched, bulk equipment and free space propagation poses limitations to scalability and real-world applications. 
	
	Consequently, in the last few years, there has been a significant focus on the generation of multi-photon states in integrated circuits, with noteworthy results in the demonstration of reconfigurable graph states in silicon-based devices via on-chip spontaneous four-wave mixing (SFWM) ~\cite{adcock2019, reimer2019, llewellyn2020, vigliar2021}. In these works, the generation of the input state is probabilistic, for it is obtained via post-selection on squeezed light generated by SPDC or SFWM. An intrinsic issue of this approach is the emission of unwanted photon pairs, whose generation probability is proportional to the average number of generated photons. One  has thus to reach a trade-off between large rate and coincidence-to-accidental ratio~\cite{takesue2010}.
	
	An alternative approach to the generation of multi-photon states harnesses optically engineered quantum-dot (QD) emitters that operate as deterministic bright sources of indistinguishable single-photons in wavelength ranges well suited for high-efficiency single-photon detectors~\cite{somaschi2016, wang2019, uppu2020, tomm2021}. Recently, the potential of such high-performance single-photon sources (SPS) for the generation of multi-photon states has been highlighted with bulk optics \cite{li2020}. QDs are also compatible with integrated photonic chips \cite{wang2019BS, de2022} and, in particular, with glass optical circuits fabricated by femtosecond laser micromachining (FLM) \cite{corrielli2021, meany2015}. These devices offer an efficient interfacing with optical fibers, low losses at the QD emission frequencies, and the possibility of integrating thermal phase shifters to achieve full circuit programmability. Thanks to these characteristics, the combined use of QD-based SPS and laser-written photonic processors have demonstrated to be an effective platform for quantum information processing \cite{anton2019, pont2022}.
	
	In this work, we demonstrate for the first time the on-chip generation and characterization of a 4-photon Greenberger-Horne-Zeilinger (GHZ) state~\cite{greenberger1989} using a bright SPS in combination with a reconfigurable laser-written photonic chip. We achieve a complete characterization of the generated state through the reconstruction of its density matrix via quantum state tomography. By using a semi device-independent approach, we further test non-classical correlations, certify entanglement, non-biseparability, and study the robustness of the generated states to noise. Finally, as a proof-of-principle that our platform is application ready, we show that it can be used to implement a 4-partite quantum secret sharing protocol ~\cite{hillery1999}. Our approach combines the practical assets of bright QD-based SPS, efficient single-photon detectors, and low-loss, scalable, integrated optical circuits fabricated using FLM.

	\section{Results}\label{sec2}
	
	\subsection{Path-encoded 4-GHZ generator}\label{sec:4GHZ}
    Among graph states, GHZ states are striking examples of maximally entangled states that are considered a pivotal resource for photonic quantum computing, since they can be used as building blocks for the construction of high-dimension cluster states \cite{li2015}. They are also of interest for quantum communication and cryptography protocols~\cite{hillery1999, proietti2021}. 
	
\begin{figure}[h]
		\centering
		\includegraphics[width=0.9\linewidth]{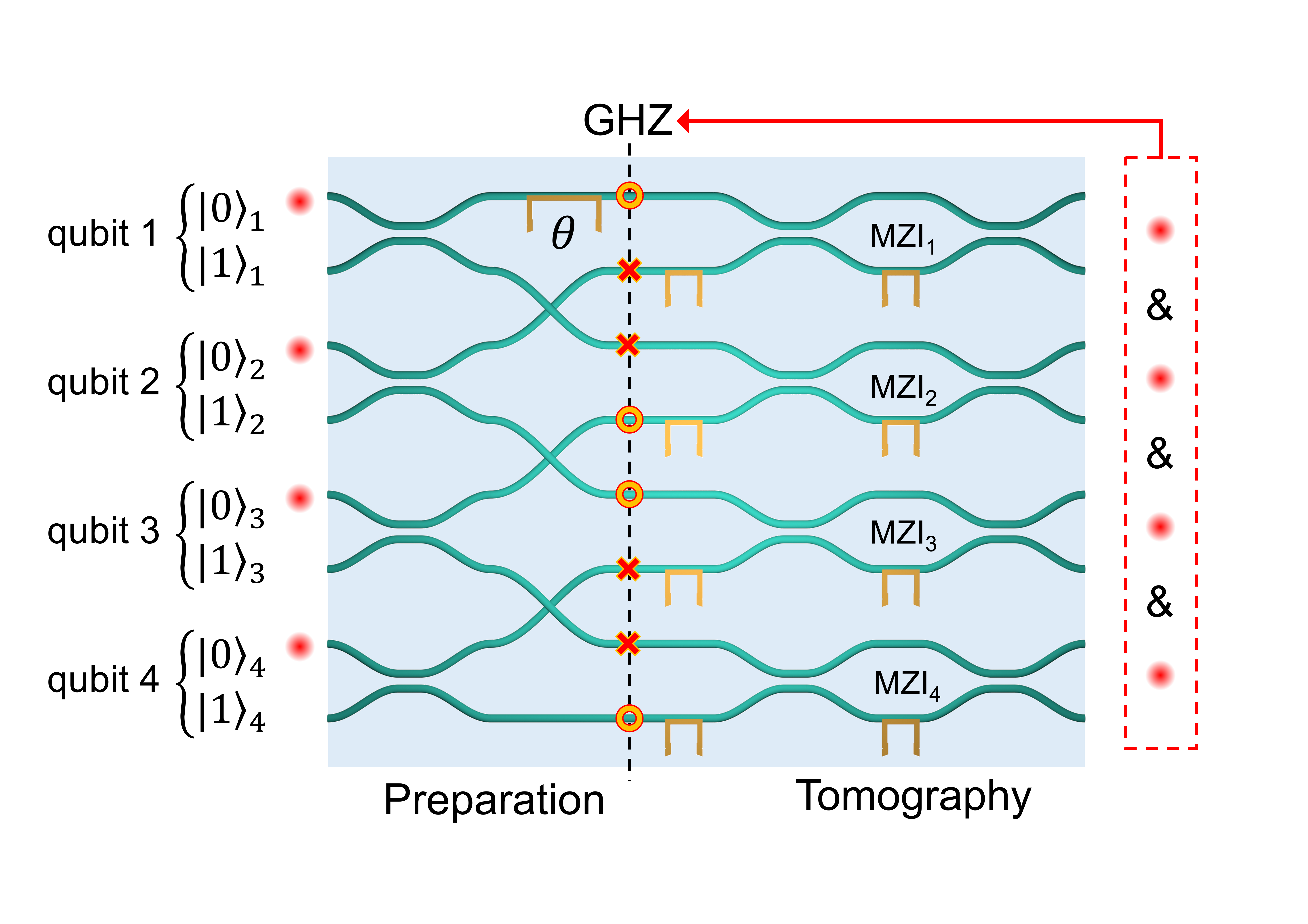}
		\caption{\textbf{Integrated path-encoded 4-GHZ generator}.
		Conceptual layout. For each qubit i, the upper and lower waveguides  encode the computational basis \{$\ket{0}_i$, $\ket{1}_i$\}. The \textbf{preparation} of the state in Eq.~\eqref{eq:4GHZ} along the orange dashed line (dots and crosses encoding the $\ket{0101}$ and $\ket{1010}$ states respectively) is conditioned (''\&'' and red arrow) to the detection of one and only one photon (red dot) per qubit. The projective measurements of the \textbf{tomography stage} are performed by four thermally reconfigurable MZIs. One of the phase shifters before MZI$_1$ is also used for controlling the phase $\theta$ (Supplementary~\ref{supp:exp}).}
		\label{fig:Fig1}
	\end{figure}

	In this work, we target 4-partite GHZ states of the form:
    \begin{equation} \label{eq:4GHZ}
    	\ket{\text{GHZ}_4}=\frac{\ket{0101}+\ket{1010}}{\sqrt{2}}
	\end{equation}

    \noindent encoded in the path degree of freedom (dual-rail). In Fig.\ref{fig:Fig1} the conceptual scheme of our path-encoded 4-partite GHZ generator chip is depicted. It is composed by a first layer of balanced beam splitters (50/50 directional couplers) followed by waveguide permutations (3D waveguide crossings). The 4-photon input states are created using a high-performance QD based single-photon source (\textit{Quandela e-Delight-LA})~\cite{somaschi2016} and a time-to-spatial demultiplexer (\textit{Quandela DMX6}) (see Methods), which initialise the input states to $\ket{0000}$. With this scheme, the generation of the GHZ states is conditioned to the presence of one and only one photon per qubit. Finally, the chip allows for the characterisation of the generated states by means of four reconfigurable Mach-Zehnder interferometers (MZI), each one implementing single-qubit Pauli projective measurements ($\sigma_x$, $\sigma_y$, $\sigma_z$) in the path degree of freedom. The overall system efficiency enables us to detect useful 4-fold coincidence events at the rate of 0.5 Hz with a pump rate of 79~MHz. This is on par with the recent record for generating entangled states in integrated photonics~\cite{vigliar2021}. Because of the short lifetime of our photons (145 ps), a pump rate of 500~MHz is achievable, which would yield a generation rate $>$3 times higher than~\cite{vigliar2021}. Further details about the chip functioning, its manufacturing and the experimental setup are provided in the Supplementary Information (\ref{supp:exp}).
	
\begin{figure*}[t]
		\includegraphics[width=0.9\linewidth]{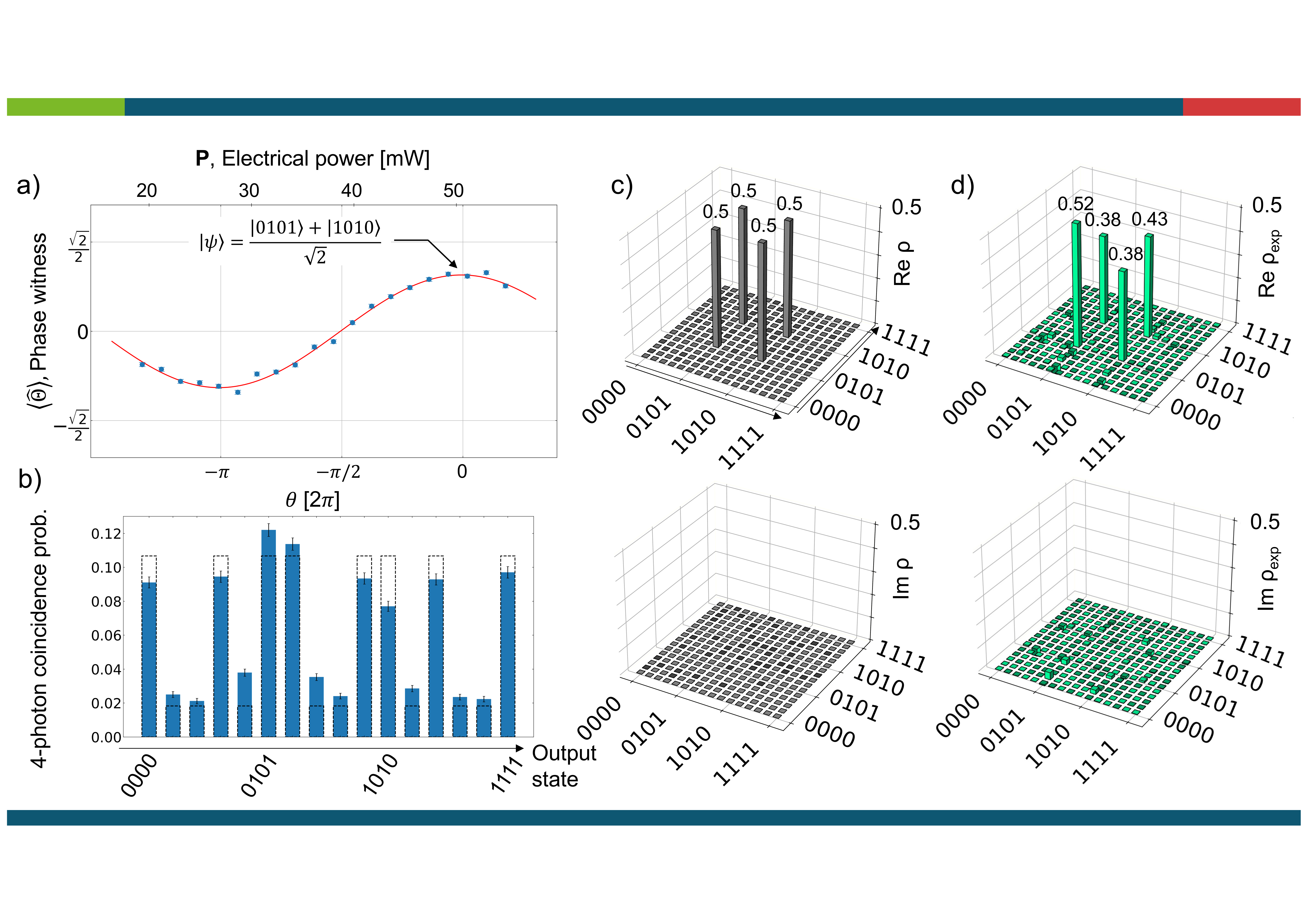}
		\caption{\textbf{Preparation of 4-GHZ states in the path-encoded basis and reconstruction of its density matrix.} (a) A single phase shifter (see Fig.~\ref{fig:Fig1}.a) is scanned over a $\sim$~50~mW range of electrical driving power P. For each value of $\theta$ and each 4-qubit state we acquire the 4-photon coincidence rates for 900~s and compute the phase witness $\langle \hat{\Theta}_{exp} \rangle$ (blue circles) which is fitted with a cosine function with free amplitude (red line). The error bars are computed assuming a shot noise limited error on the detected 4-fold coincidences. Using the fit parameters, the internal phase of the GHZ state is set to $\theta=0~[2\pi]$ when P=52.81~mW. (b) The experimental (theoretical) 4-photon coincidence probability distribution measured in this configuration are reported in blue bars (dotted black bars). Along the horizontal arrow, the qubit states are ordered as \{0000, 0001, 0010, 0011, 0100, 0101, 0110, 0111, 1000, 1001, 1010, 1011, 1100, 1101, 1110, 1111\}. (c-d) The real (top) and imaginary (bottom) parts of the reconstructed density matrix $\rho=\ket{\text{GHZ}_4}\bra{\text{GHZ}_4}$ for the target (c, grey) are compared to the reconstruction from the experimental 4-photon tomography data using maximum likelihood estimation $\rho_{exp}$ (d, green). The noise (±1e-9) in (c) arises from the numerical method (maximum likelihood estimation) and is orders of magnitude smaller than the noise arising from the imperfections of the experimental setup (±0.03) in $\rho_{exp}$.}
		\label{fig:Fig2}	
	\end{figure*}

    Our photonic chip is reconfigurable, thus it can generate a whole class of GHZ states of the form $\ket{\text{GHZ}_4}^{(\theta)}= \left( \ket{0101}+e^{i\theta}\ket{1010} \right)/\sqrt{2}$, parametrized by an internal phase $\theta$ corresponding to the algebraic sum of the optical phases acquired by each photon in the different paths of the preparation stage after the beam splitters (Supplementary \ref{supp:photoniccircuit}). The phase $\theta$ can be controlled with a single phase-shifter localized on one of these paths, as depicted in Fig.\ref{fig:Fig1}. 
    To prepare the targeted GHZ states, we use a 2-qubit Pauli projector $M_0^{(i)}$ for each qubit $i\in\{1,..,4\}$ (See Methods for each $M_0^{(i)}$ definition), and compute the expectation value $\langle \hat{\Theta} \rangle = \langle M_0^{(1)} M_0^{(2)} M_0^{(3)} M_0^{(4)} \rangle$. For our class of GHZ states,  $\langle \hat{\Theta} \rangle$ can be used as an internal phase witness as $\langle \hat{\Theta} \rangle = (\sqrt{2}/2)\cos{\theta}$, and it reaches its maximum value for the target stsate $\ket{\text{GHZ}_4}$ at $\theta=0~[2\pi]$.
    Fig.\ref{fig:Fig2}.a shows the measured $\langle \hat{\Theta}_{exp} \rangle$ as a function of the driving power of one of the outer thermal phase shifters of MZI$_1$, and it demonstrates that we have full control over the value of $\theta$ in the state preparation. The recorded 4-photon coincidence probability distribution corresponding to our target state is reported in Fig.~\ref{fig:Fig2}.b. We measure the maximal value of $\langle \hat{\Theta}_{exp} \rangle=0.56\pm0.01$, which is limited by some experimental imperfections discussed in Sec.\ref{sec:Tomography}.
    
	There are many tools available to detect entanglement and to estimate the fidelity of a multipartite system with minimal resources. Entanglement witnesses~\cite{guhne2007, guhne2009}, Bell-like inequalities~\cite{Baccari2017}, or the so-called GHZ paradox~\cite{greenberger1989, mermin1990} all require only a few Pauli projective measurements to characterize the GHZ states. Here we choose the stabilizer witness for GHZ states $\mathcal{W}_{\text{GHZ}_4}$ (Methods) first introduced in \cite{toth2005}, for which a measured negative expectation value signals the presence of entanglement. This witness requires only two projective measurements to detect entanglement, and to compute a lower bound of the fidelity of the generated state to our target $\ket{\text{GHZ}_4}$, where $\mathcal{F}_{\text{GHZ}_4} \ge (1-\langle\mathcal{W}_{\text{GHZ}_4}\rangle)/2$. 
	We found $\langle \mathcal{W}_{\text{exp}} \rangle=-0.65\pm0.03$,
	which certifies that the generated state is entangled. The lower bound for the fidelity of the generated state to the target is $\mathcal{F}_{\text{GHZ}_4}\ge0.82\pm0.01$.

	\subsection{On-chip quantum-state tomography}\label{sec:Tomography}
	
	The generated states can be fully characterised through the reconstruction of the density matrix via maximum likelihood estimation~\cite{altepeter2005} from a full quantum-state tomography. Most previous on-chip entanglement generation protocols based on SPDC or SFWM sources use partial analysis of the state, such as an entanglement witness or the stabilizer group, to determine the state fidelity with respect to the target and to detect entanglement. Here the high single-photon rate of the QD source and the low insertion losses of the chip allow us to obtain the density matrix of the 4-qubit state.
	\begin{table*}
        \includegraphics[width=\linewidth]{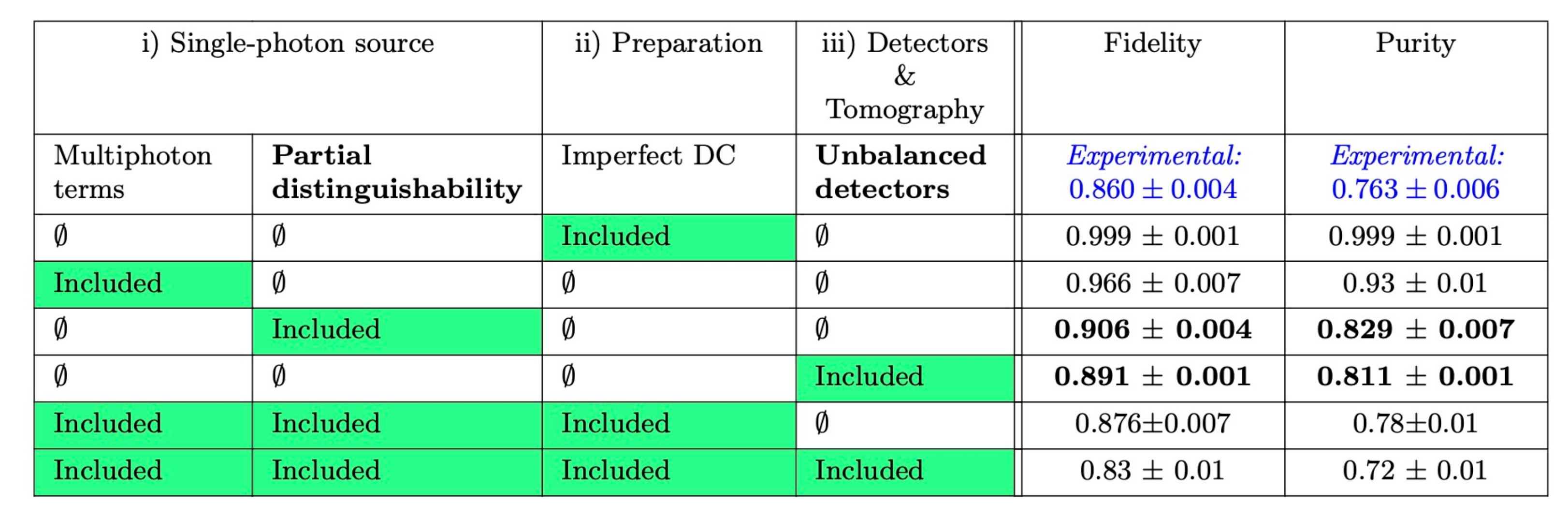}
        \caption{Numerical simulation of the fidelity and the purity of a 4-qubit GHZ state computed from the reconstruction of the density matrix with maximum likelihood estimation~\cite{simulation2022}. The impact of each dominant experimental source of noise (second row) is assessed using independently measured parameters (Supplementary~\ref{Simulation}). The main source of noise (bold) are the partial distinguishability of the input 4-photon states and the unbalanced efficiency of the single-photon detectors. Error bars are obtained from Monte Carlo simulations assuming a Poissonian distribution of the measured counts when measuring the sources of noise.}
        \label{tab:simu}
    \end{table*}
	
	To fully reconstruct the $16\times16$ density matrix, $3^4=81$ projective measurements, corresponding to all possible combinations of $(\sigma_x,\sigma_y,\sigma_z)$ among the four qubits, are necessary. The density matrix is determined by using a maximum likelihood estimation to restrict the numerical approximation to physical states. The result is shown in Fig.\ref{fig:Fig2}.d. From the experimental density matrix $\rho_{\text{exp}}$ we calculated the fidelity $\mathcal{F}_{\text{GHZ}_4}=\bra{\text{GHZ}_4}\rho_{exp}\ket{\text{GHZ}_4}=0.860\pm0.004$ and the state purity $\mathcal{P}_{\text{GHZ}_4}=Tr\{ \rho_{exp}^2 \}=0.763\pm0.006$. Our four-photon results establish a new state-of-the-art in terms of fidelity and purity for integrated implementations of GHZ states. Previous record values demonstrated in Ref.~\cite{vigliar2021} showed fidelity of $0.80\pm0.02$  and a purity of $0.72\pm0.04$ which was achieved for a two-photon GHZ and not a multipartite state as in our case.
    
	In what follows we investigate all the sources of noise in our system to analyse quantitatively what is limiting our values of fidelity and purity. In order to explore the effect of each experimental imperfection, we use a phenomenological model (Supplementary~\ref{Simulation}) based on the measured characteristics of the experimental setup to perform numerical simulations of the experiment~\cite{simulation2022}. The model accounts for 
	i) the imperfections of the single-photon source (Supplementary~\ref{imperfect sps}), namely the imperfect single-photon purity and indistinguishability of the input state made of four simultaneous photons, 
	ii) the imperfections of the preparation of the GHZ states (Supplementary~\ref{imperfect chip}), namely the imperfect directional couplers and initialisation of the internal phase $\theta$ in the preparation stage (see Fig.~\ref{fig:Fig1}), and 
	iii) the imperfections of the projective measurements, implemented by the MZIs,  and detectors (Supplementary~\ref{supp:projectivemeasurements}) experimentally dominated by unbalanced detection efficiencies, modeled by imperfect projective measurements. Each imperfection is studied independently to uncover the main source of noise in the system. The results of the numerical simulations and the corresponding values of fidelity and purity are shown in Tab~\ref{tab:simu}.
	
	The imperfections of the single photon source, namely the multiphoton terms and the partial distinguishability of the 4-photon input state, limit the achievable values of fidelity and purity. The multiphoton component $g^{(2)}(0)$ of the single-photon stream, measured independently in a Hanbury-Brown and Twiss setup, is $g^{(2)}(0)=0.005\pm0.001$. The indistinguishability of two subsequent photons (12.3~ns time delay) measured with a Hong-Ou-Mandel (HOM) interferometer~\cite{ollivier2021} is $M_s=0.962\pm0.02$. The indistinguishability of the 4-photon input state is limited by long-term fluctuations of the emitter environment (electrical and magnetic noise)  when using the time-to-spatial demultiplexing scheme that synchronizes photons up to 500~ns apart (see setup in \ref{supp:opticalsetup}). The indistinguishability of long-delay (500~ns time delay) photons is measured through the demultiplexer and the chip used as multiple HOM interferometers (Supplementary~\ref{HOM}). We measure a minimal 2-photon interference visibility $V_{\text{min}}=0.88\pm0.01$. The indistinguishability of the photons is degraded by the imperfect temporal overlap and imperfect polarisation control of the photons at the output of the demultiplexer. It is also affected by fabrication imperfections of the optical circuit. We use all the accesible pairwise indistinguishabilities (see Supplementary~\ref{HOM}) as inputs for the model. All the imperfections from the chip input to the detectors are thus taken into account twice, which explains why we underestimate the fidelity and purity when all the sources of noise are taken into account.

    \subsection{Causal inequality for GHZ state certification}
    \label{sec:Bell}

     \begin{figure*}[t]
        \centering
       	\includegraphics[width=0.8\linewidth]{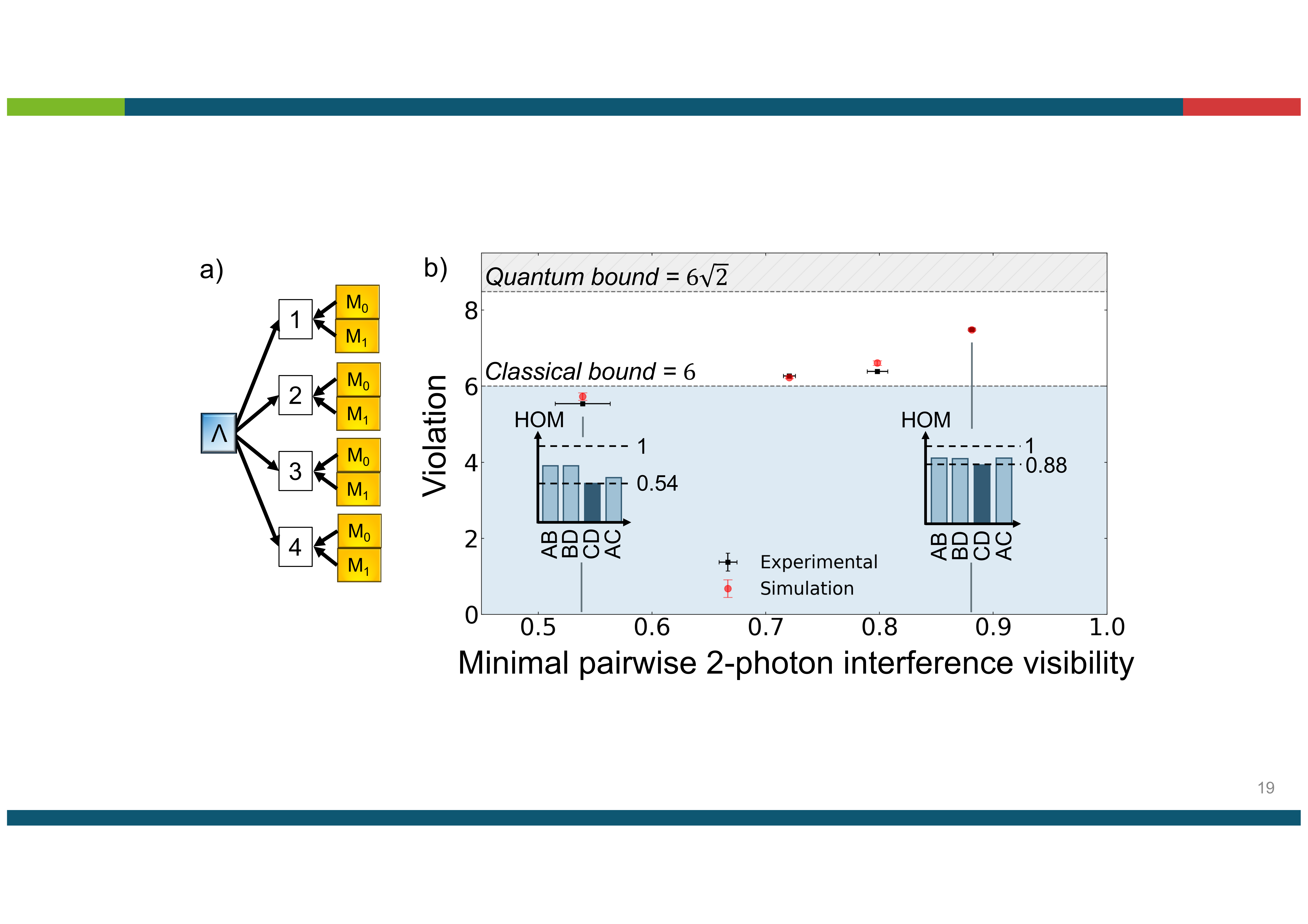}
    	\caption{\textbf{Semi-device independent certification of the GHZ$_4$ state} \textbf{(a) Causal structure.} In the DAG shown here, there are three different kinds of nodes: $\Lambda$ is a hidden variable (blue box), (M$_0$, M$_1$) are measurement settings (yellow boxes) and each party is associated with a variable (i) outputting measurement outcomes. \textbf{ (b) Violation of the Bell-like inequality characterizing 4-partite GHZ states.} The violation of the Bell-like inequality certifies the presence of non-classical correlations. We study the dependence of the entanglement of the GHZ state on the dominant source of noise, i.e. the minimal pairwise 2-photon indistinguishability. A half-wave plate is used to rotate the polarisation of photon C and make it distinguishable from A, B and D. The measured 2-photon mean wavepacket overlap (HOM) is shown in the insets for 2 datapoints. The experimental data (black squares) and simulation (red circles) demonstrates a good numerical match. Error bars for the violation of the inequality, referring to the 95\% confidence interval, are obtained from error propagation assuming a shot noise limited error on the total number of 4-fold coincidences. Error bars for the minimal HOM correspond to the standard deviation of the measured 2-photon interference visibility distribution measured between each combination of binary measurements. The violation of the Bell-like inequality reaches a maximal value of    $\mathcal{I}^2=7.49\pm0.04$ when all photon have the highest pairwise indistinguishability, which certifies non-classical correlations an non bi-separability within 39 standard deviations.}
    	\label{fig:Fig3}
    \end{figure*}

    We further certify the presence of non-classical correlations within the generated state, by adopting an approach requiring fewer measurements than the full quantum state tomography and minimal assumptions on the experimental apparatus, i.e in a semi device-independent fashion.

    We use Eq.~\eqref{eq:ine} as a special case of generic Bell Inequalities for self-testing graph states that can be found in \cite{baccari2020}.
    Under the assumptions detailed and justified in Supplementary~\ref{supp:BLI}, a violation of Eq.~\eqref{eq:ine} guarantees the presence of non-classical correlations among the parties. Like for a two-partite Bell measurement, the orthogonal measurement bases $(M_0^{(i)}, M_1^{(i)}) $ for each of the four qubits $i \in \{1,..,4\}$ have been set to obtain the highest violation of Eq.~\eqref{eq:ine} for the target state $\ket{\text{GHZ}_4}$, i.e. $6\sqrt{2}=8.48$. Each of these 2-qubit Pauli projectors are defined in the Methods section. A maximal quantum violation self-tests that the generated state has the form of the target state~\cite{agrestiprx,PhysRevLett.127.230503}.
    
    \begin{widetext}
        \begin{equation}
        \begin{split}
        	\mathcal{I}^2= \sum_{i=2}^4\langle M_0^{(1)}M_1^{(i)}\rangle-\sum_{i=2}^4 \langle M_1^{(1)}M_1^{(i)} \rangle + 3 \langle M_0^{(1)}M_0^{(2)}M_0^{(3)}M_0^{(4)}\rangle + 3 \langle M_1^{(1)}M_0^{(2)}M_0^{(3)}M_0^{(4)}\rangle \leq 6
        	\label{eq:ine}
        \end{split}
        \end{equation}
    \end{widetext}

    We compute the expectation values of the left hand side of Eq.~\eqref{eq:ine} (see Methods). Abiding Eq.~\eqref{eq:ine} would mean the measured probabilities could be compatible with a local hidden variable model, according to the directed acyclic graph in Fig.\ref{fig:Fig3}.a~\cite{pearl}. On the contrary, a violation of Eq.~\eqref{eq:ine} certifies the presence of non-classical correlations among the parties. In our case, the largest experimental estimate of $\mathcal{I}^2$ is $7.49 \pm 0.04 > 6$, which violates the classical bound described in Eq.~\eqref{eq:ine} by 39 standard deviations.

    Furthermore, we address the robustness of our inequality violation with respect to the experimental noise. Since we have identified partial distinguishability as the main source of noise in our system, we vary in a controlled way the indistinguishability among the parties, i.e. make one of the four photons distinguishable in the polarisation degree of freedom using a half-wave plate, to gauge how robust the entanglement is with respect to this issue. In Fig.~\ref{fig:Fig3}.b, we report the measured values for $\mathcal{I}^2$ while increasing the photon distinguishability, which we calibrated with independent measurements (Supplementary~\ref{HOM}). Our setup can tolerate a substantial amount of distinguishability before inequality $\mathcal{I}^2$ is not violated anymore. In Fig.~\ref{fig:Fig3} we observe a good agreement between experimental data and simulations.b for different levels of noise, which reveals that our model can faithfully describe the generated state for a wide range of input parameters.  

	\subsection{Quantum secret sharing}\label{sec:QSS}
	
	We now examine the suitability of our approach to implement "Quantum secret sharing" (QSS) -- a protocol presented in 1999 by Hillery \textit{et al.}~\cite{hillery1999}. QSS considers the practical case of a regulator who wants to share a secret string of random bits with three interlocutors, in such a way that they can access the secret message only if they cooperate all together. {In this protocol, the regulator prepares a string of 4-qubits in a state of the form of Eq.~\eqref{eq:4GHZ}, keeps one qubit  and distributes the three others to three parties.  All four parties then randomly choose a basis for measuring the state of their qubit: $\sigma_x$ or $\sigma_y$.  The sifted key is extracted, on average with a 50\% success probability, after public basis sharing (see Supplementary~\ref{QSS} for more details).}
	
	\noindent We performed a proof-of-principle implementation of this QSS protocol by generating the 4-qubit GHZ state with our chip and by exploiting the reconfigurable MZIs to perform the required projective measurements.
	Each party measures its share of the 4-qubit GHZ state by randomly selecting a measurement basis, and by recording the measurement outcome in the raw key when the first 4-photon coincidence event occurs. This procedure is repeated until the target length for the raw key has been reached. The key sifting is then performed by discarding the raw bits that correspond to non-valid basis choices. The raw bit generation rate we obtained is about 0.5~Hz. This rate incorporates the dwell time to reach stable settings ($\sim$ 100~ms) for the randomly chosen projective measurements.
	
	We evaluate the total number of errors by calculating the quantum bit error rate (QBER) on the sifted key. The most accurate uninterrupted run has a QBER of 10.87\% $\pm$ 0.01, which guarantees a secure communication as it is below the required threshold of 11\%~\cite{shor2000}, with a raw length of 4060 bits, and a sifted length of 1978 bits.

	\section{Discussion}
	
    In this work we demonstrated the generation of 4-photon multipartite GHZ states integrating a deterministic solid-state QD-based SPS with a reconfigurable glass photonic chip. We achieved a  post-selected 4-fold coincidence rate of 0.5~Hz with a pump rate of 79~MHz, which allowed us to perform full quantum state tomography, with a fidelity of 86\% to the target state. A  4-photon coincidence rate of 10~Hz was reached removing one stage of spectral filtering on the single photon source  with a limited effect ($+0.007$ on $g^{(2)}(0)$ and $-0.05$ on $M_{s}$, see Section~\ref{supp:opticalsetup}). The combination of high-fidelity and high rate, as well as the overall platform stability ---stable enough for highly demanding  measurements such as full quantum state tomography--- shows the suitability of the platform for intermediate scale computing or communication protocols.\\
    
    Our systematic analysis and numerical modeling shows that, despite state of the art performances of the source,  the effective overall indistinguishability of the photons is the main limiting factor to the ideal fidelity.
   We thus identify many handles to further improve these results. Our observations indeed indicate that the effective indistinguishability of the QD photons at long delays is limited by the stability of the voltage source used to operate the QD. 
    More precise temporal overlap of the photons and better polarization control at the input of the chip would also increase the net photon overlap. In the current demonstration, we have 50~\% insertion losses on the chip. We foresee that this can be improved to 30~\%. Finally, the operation rate can be brought up to at least to 500 MHz thanks to the short photon profile. All these improvements would allow to significantly improve state fidelity and purity, and increase performances for applications, i.e. a higher bitrate and a lower QBER in the QSS protocol.\\
    
    All together, these qualitative and quantitative results constitute an important milestone in the generation and use of high-dimensional quantum states. They prove that the integration of QD-based SPSs with glass chips for generation and manipulation of multipartite states is now mature and can lead to performances comparable to those achievable by state-of-the-art free space implementations. The deterministic nature of QD-based SPSs along with the integration of low-loss, stable, and re-configurable optical elements in a scalable photonic glass chip put our platform in the front line for the development of practical photon-based quantum devices.\\
    Recent works have shown the ability to generate linear cluster state at high rate using an entangling-gate in a fiber loop~\cite{istrati2020}, or through spin-photon entanglement~\cite{coste2022}, harnessing a similar QD-based single-photon source. Generating multipartite entanglement on chip, as demonstrated here, will be key to obtain the 2D cluster states required for measurement-based quantum computation.
	
	\section*{Acknowledgements}
	
	This work is partly supported the European Union's Horizon 2020 FET OPEN project PHOQUSING (Grant ID 899544), the European Union's Horizon 2020 Research and Innovation Programme QUDOT-TECH under the Marie Sklodowska-Curie Grant Agreement No. 861097,  the French RENATECH network, the Paris Ile-de-France Region in the framework of DIM SIRTEQ. The fabrication of the photonic chip was partially performed at PoliFAB, the micro- and nanofabrication facility of Politecnico di Milano (www.polifab.polimi.it). F.C. and R.O. would like to thank the PoliFAB staff for the valuable technical support.
	
	\section*{Author contributions}
	
		The development of the experimental platform used in this work was made possible by the collaboration of multiples teams, as revealed by the large number of authors and institutions involved. M.P.: Single-photon source (SPS) sample design, experimental investigation, data analysis, formal analysis, numerical simulations, methodology, visualization, writing, G.Co. conceptualization, methodology, photonic chip fabrication \& characterisation, visualization, writing, A.F.: experimental investigation, data analysis, numerical simulations, I.A.: data analysis, numerical simulations, visualization, writing, G.Ca.: conceptualization, formal analysis, visualization, writing N.M.: experimental investigation, supervision, P-E.E.: conceptualization, formal analysis, writing, F.C., R.A., P.H.D.F.: photonic chip fabrication \& characterisation, N.S., I.S.: SPS nano-processing, J.S.: numerical simulations, M.M., A.L.: SPS sample growth, P.S.: SPS sample design \& nano-processing, conceptualization, methodology, supervision, writing, funding acquisition, F.S.: conceptualization, supervision, funding acquisition M.L.: conceptualization, methodology, visualization, writing, N.B, R.O: conceptualization, methodology, data analysis, visualization, writing, supervision, funding acquisition.

	\section*{Data \& code availability}
	
	The data generated as part of this work is available upon reasonable request (mathias.pont@polytechnique.org). The code used for the numerical simulations is available at~\cite{simulation2022}.
	
	\bibliographystyle{apsrev4-1}
	\bibliography{bibliography.bib}
	
	\newpage
	
	\section*{Methods}
	
	\subsection*{Single-photon source}
	The bright SPS consists of a single InAs QD deterministically embedded in the center of a micropillar~\cite{somaschi2016}. The sample was fabricated using the in-situ fabrication technology~\cite{dousse2008, nowak2014} from a wafer grown by molecular beam epitaxy composed of a $\lambda$-cavity and two distributed Bragg reflectors (DBR) made of GaAs/Al$_{0.95}$Ga$_{0.05}$As $\lambda$/4 layers with 36 (18) pairs for the bottom (top). The top (bottom) DBR is gradually p(n)-doped and electrically contacted. The resulting p-i-n diode is driven in the reversed bias regime to reduce the charge noise~\cite{berthelot2006} and to tune the QD in resonance with the microcavity. The resonance of the QD with the cavity mode at $\lambda_{\text{QD}}$=928~nm is actively stabilised in real time with a feedback loop on the total detected single-photons countrate. The sample is placed in a closed-loop cryostat operating at 5~K. The LA phonon-assisted excitation~\cite{thomas2021} is provided by a shaped Ti:Sa laser at $\lambda_{\text{excitation}}$=927.4~nm, generating $\sim15$~ps pulses with a repetition rate of 79 MHz. The (polarised) first-lens brightness, defined as the (polarised) single-photon countrate before the first optical element computed from the loss-budget presented in Supplementary~\ref{Losses} is ($\beta_{FL} \sim 38\%$) $\beta_{FL} \sim 50\%$, leading to a detected countrate of 18.9~MHz (12.3~MHz) accounting (not accounting) for the 65\% efficiency of the SNSPD. To improve the single-photon purity and indistinguishability of the source a narrow optical filter ($\text{FWHM}_{\text{filter}}=4\times\text{FWHM}_{\text{photon}}$, T=60\%) is added to the laser filtering module (Supplementary~\ref{supp:opticalsetup}). With this additional spectral filter the detected single-photon countrate is 11.3~MHz (7.4~MHz).
	
	The single-photon stream is split into four spatial modes using an acousto-optic based time-to-spatial demultiplexer. The time of arrival of each photon at the input of the optical circuit is synchronised with fibered delays (0~ns, 180~ns, 360~ns, 540~ns). The polarisation of each output is actively controlled with motorised paddles for five minutes every one hour, to account for the temperature instability in the laboratory.
	
	\subsection*{Projectors}
	
	The projectors $M_0^{(i)}$ and $M_1^{(i)}$ used in the definition of the operator $\hat{\Theta}$ for the characterization of the phase $\theta$, and in the Bell-like inequality expressed by Eq.~\eqref{eq:ine} are: $M_0^{(1)}= \frac{\sigma_x+ \sigma_z}{\sqrt{2}}$, $M_1^{(1)}= \frac{\sigma_x-\sigma_z}{\sqrt{2}}$, $M_0^{(3)}= \sigma_x$, $M_1^{(3)}= \sigma_z$, $M_0^{(2)}=M_0^{(4)}= -\sigma_x$ and $M_1^{(2)}= M_1^{(4)}=-\sigma_z$.
  
	\subsection*{Expectation values}
	
	For a given 4-qubit projector $\hat{E}$, the expectation value is computed as $\langle \hat{E} \rangle=\sum_{i=1} ^{16} p_i \mathcal{E}_i$, where $p_i$ is the probability of detecting the 4-qubit output state $i$ associated to the measured normalized 4-photon coincidence rate of each possible output state and $\mathcal{E}_i=\pm1$ is the product of the individual outcomes where +1 is associated with the detection of $\ket{0}$ and -1 is associated with $\ket{1}$. Note that whenever we have $\mathds{1}$ in an expectation value for a qubit then we always record +1 (irrespective of which detector have clicked) which amounts to trace out the corresponding qubit.
	
	\subsection*{Stabilizer witness}
	
	The stabilizer witness $\mathcal{W}_{\text{GHZ}_4}$~\cite{guhne2007} can be computed from the generating operators $g_1^{(\text{GHZ}_4)}=\sigma_x\otimes\sigma_x\otimes\sigma_x\otimes\sigma_x=\sigma_x^{\otimes4}$ where $\otimes$ is the Kronecker product of the Pauli matrices, and $g_k^{(\text{GHZ}_4)}=-\sigma_z^{(k-1)}\otimes\sigma_z^{(k)}$ for $k=2,3,4$ (the identity operator $\mathds{1}$ has been omitted for two of the parties not involved) as
	
	\begin{equation} \label{eq:WGHZ}
	    \frac{\mathcal{W}_{\text{GHZ}_4}}{3}=\mathbb{1}-\frac 2 3 \left[ \frac{g_1^{(\text{GHZ}_4)}+\mathbb{1}}{2}+\prod_{k=2}^4 \frac{g_k^{(\text{GHZ}_4)}+\mathbb{1}}{2} \right].
	\end{equation}
	
	 To compute these expectation values, we need to perform only two projective measurements, namely $\sigma_x^{\otimes4}$ and $\sigma_z^{\otimes4}$. This witness allows to give a lower bound on the fidelity via $\mathcal{F}_{\text{GHZ}_4} \ge (1-\langle\mathcal{W}_{\text{GHZ}_4}\rangle)/2$. A measured negative expectation value signals the presence of entanglement. 
	
	\clearpage
	
	\pagebreak
	\widetext
	\begin{center}
		\textbf{\large Supplementary Information}
	\end{center}
	
	\setcounter{section}{0}
	\setcounter{equation}{0}
	\setcounter{figure}{0}
	\setcounter{table}{0}
	\setcounter{page}{1}
	\makeatletter
	\renewcommand{\thetable}{S\arabic{table}}
	\renewcommand{\thesection}{S-\Roman{section}}
	\renewcommand{\theequation}{S\arabic{equation}}
	\renewcommand{\thefigure}{S\arabic{figure}}
	
	\section{Experimental implementation}\label{supp:exp}
	
	\subsection{The photonic circuit}\label{supp:photoniccircuit}
	\begin{figure}[h]
		\centering
		\includegraphics[width=0.9\linewidth]{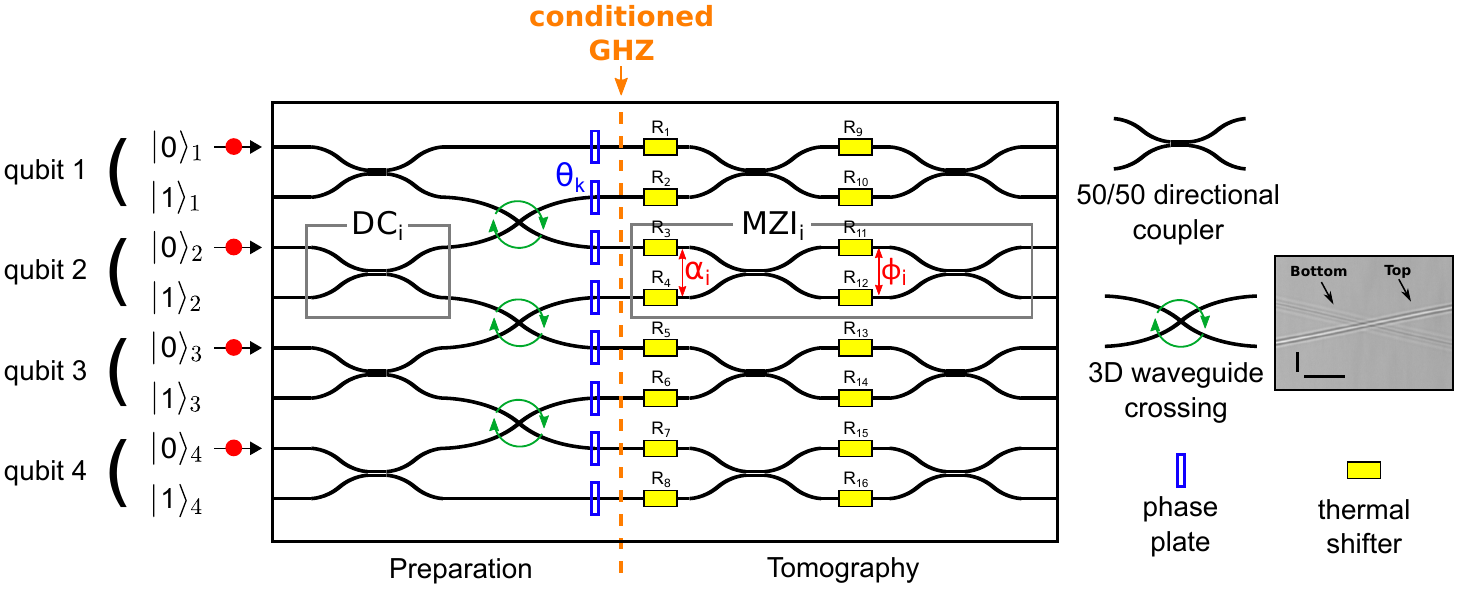}
		\caption{(a) Schematic layout of the integrated photonic chip for 4-photon path-encoded GHZ state generation. In the right inset a microscope picture of a waveguide crossing is shown. Inset scalebars correspond to 150 $\mu$m (horizontal) and 15 $\mu$m (vertical).
		}
		\label{fig:chip}
	\end{figure}

	In Fig. \ref{fig:chip} the schematic of the chip functioning is depicted. It's layout is inspired from Bergamasco \textit{et al.}~\cite{bergamasco2017}, where a scheme for the generation of a path-encoded 3-qubit GHZ state is presented. Here, the scheme is generalized to 4 qubits.\\
	Up to a global phase with no physical meaning, the field scattering matrix of this device (with spatial modes numbered top to bottom from 1 to 8) from input to the orange dashed line reads as:
	\begin{equation}
	    \label{eq:scattering}
	    U=\frac{1}{\sqrt{2}}
	    \begin{bmatrix}
	        e^{i\theta_1} & ie^{i\theta_1} & 0 & 0 & 0 & 0 & 0 & 0  \\
	        0 & 0 & e^{i\theta_2} & ie^{i\theta_2} & 0 & 0 & 0 & 0 \\
	        ie^{i\theta_3} & e^{i\theta_3} & 0 & 0 & 0 & 0 & 0 & 0 \\
	        0 & 0 & 0 & 0 & e^{i\theta_4} & ie^{i\theta_4} & 0 & 0 \\
	        0 & 0 & ie^{i\theta_5} & e^{i\theta_5} & 0 & 0 & 0 & 0 \\
	        0 & 0 & 0 & 0 & 0 & 0 & e^{i\theta_6} & ie^{i\theta_6} \\
	        0 & 0 & 0 & 0 & ie^{i\theta_7} & e^{i\theta_7} & 0 & 0 \\
	        0 & 0 & 0 & 0 & 0 & 0 & ie^{i\theta_8} & e^{i\theta_8} \\
	    \end{bmatrix}.
	\end{equation} 
	The values $\theta_k$ indicate the optical phase that ligth acquires in propagating from the chip input to the orange dashed line in correspondence of spatial mode $k$, and are represented as static phase plates in Fig.~\ref{fig:chip}.
	Consecutive pairs of odd and even spatial modes are used to encode the qubit computational basis states $\ket{0}$ and $\ket{1}$. If the circuit is fed with four indistinguishable photons in the separable state $\ket{\Psi_{\text{in}}}=\ket{0000}$, it is possible to show, from equation \ref{eq:scattering} and applying the standard rules to compute the output state of a linear optical network when single photons are used as input \cite{crespi2015suppression}, that, up to a global phase term, the qubit state at the orange dashed line, conditioned to the presence of one and only one photon per qubit, is a 4-fold GHZ state of the form:
	\begin{equation} \label{eq:ghz}
		\ket{\text{GHZ}_4}^{(\theta)}=\frac{\ket{0101}+e^{i\theta}\ket{1010}}{\sqrt{2}}.
	\end{equation}
	This process takes place with probability 1/8. The phase term $\theta$ within equation \ref{eq:ghz} reads as:
	\begin{equation}\label{eq:phaseGHZ}
	    \theta=\theta_1-\theta_2-\theta_3+\theta_4+\theta_5-\theta_6-\theta_7+\theta_8
	\end{equation}
	After the dashed line, a set of four reconfigurable Mach Zehnder interferometers (MZIs) allows to perform the projective measurements (Supplementary~\ref{supp:BLI}) required to characterize the generated state.

	\begin{figure}[h]
		\centering
		\includegraphics[width=0.9\linewidth]{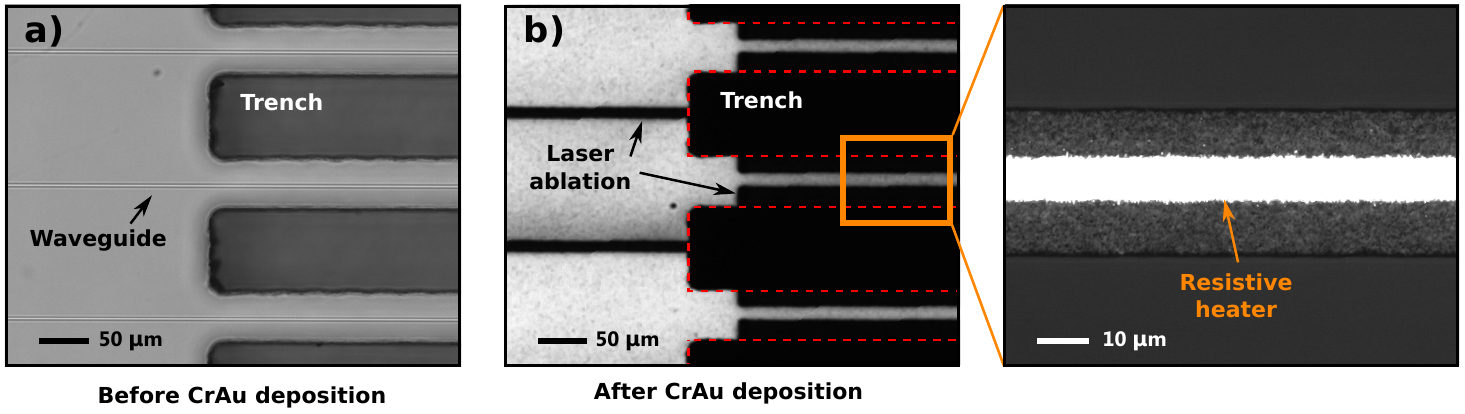}
		\caption{a) Microscope picture of the photonic chip after the fabrication of the isolation trenches around the waveguide in the thermal-shifter region. b) Microscope picture of the same view of panel a) after the deposition of the CrAu layer and the laser ablation of the resistive circuit. Bright regions correspond to CrAu coated surfaces.}		
		\label{fig:shifters}
	\end{figure}
	The optical circuit was fabricated using femtosecond laser micromachining technology~\cite{meany2015, corrielli2021} on a borosilicate glass substrate (EagleXG), by using a commercial femtosecond laser source emitting infrared ($\lambda$=1030 nm) laser pulses with duration of 170 fs and at the repetition rate of 1 MHz. The focusing optics used was a 0.5 numerical aperture water-immersion microscope objective. Single-mode waveguides for 926 nm light have been obtained with 320 nJ/pulse energy, a writing velocity of 20 mm/s, and by performing 6 overlapped writing scans per waveguide The inscription depth is 35 $\mu$m below top surface. After laser irradiation, a thermal annealing treatment (same recipe as \cite{piacentini2021}) was also performed for improving waveguide performances.
    The overall circuit length is 4 cm. All directional couplers (DC) are identical and their geometry is optimized to reach balanced 50/50 splitting ratio. The actual values of the DC reflectivities (defined as the fraction of light power that remains in the copler BAR mode) have been experimentally characterized with a 926 nm laser diode for horizontally (H) and vertically (V) polarized ligth, and the results are reported in Tab.~\ref{tab:DC}.\\
    The waveguide crossings are implemented by spline trajectories, and the relative waveguides distance at the crossing point is 15 $\mu$m, resulting in an optical cross talk $<$~-50~dB (no cross talk detected, upper bound set by the resolution of the measurement).\\
    For ensuring full MZIs programmability, a (redundant) number of 16 thermal phase shifters have been integrated in the chip, distributed as depicted in Fig.~\ref{fig:chip}. The thermal phase shifters are fabricated according to the geometry presented in ref. \cite{ceccarelli2020low} for improving the power consumption and thermal cross talk between adjacent shifters. In particular, waveguide isolation trenches (100 $\mu$m wide) are fabricated by water-assisted laser ablation between neighbouring waveguides (separated by 127 $\mu$m), in correspondence to the position of the thermal phase shifters (Fig.~\ref{fig:shifters}.a). Then, a Cr-Au metallic bilayer (5 nm Cr + 100 nm Au) is deposited on the top surface of the sample by thermal evaporation and annealed at 300 °C for 3~h in vacuum. Finally, the resistive heaters (length of 3 mm, width of 10 $\mu$m) and the corresponding contact pads are patterned by fs-laser irradiation (see Fig.~\ref{fig:shifters}.b). The electrical resistances of all thermal shifters in the chip fall within the range 410 $\Omega$ - 430 $\Omega$. Finally, the photonic chip is permanently glued to optical fiber arrays (fiber model: FUD-3602, from Nufern) at both input and output, and the overall fiber-to-fiber device insertion losses are $\sim$2.7~dB for all channels.\\
	
	Due to the presence of thermal cross talk between thermal shifters belonging to the same columns, the relations that link the currents $I_j$ dissipated on the resistors R$_j$ and the induced phase shifts $\alpha_i$ and $\phi_i$ in the MZIs are well approximated by:
	\begin{align}
	    [\alpha_1~\alpha_2~\alpha_3~\alpha_4]^T&=A~[I_1^2~I_2^2~I_3^2~I_4^2~I_5^2~I_6^2~I_7^2~I_8^2~]^T,\\
	    [\phi_1~\phi_2~\phi_3~\phi_4]^T&=\Phi_0+B~[I_9^2~I_{10}^2~I_{11}^2~I_{12}^2~I_{13}^2~I_{14}^2~I_{15}^2~I_{16}^2]^T,
	\end{align}
	where:
	\begin{align}
	    A &= 
	    \begin{bmatrix}
	        53.031 & -54.123 & -10.807 & -4.293 & -2.302 & -1.307 & -1.000 & -0.733\\
            2.915 & 9.016 & 49.504 & -48.858 & -9.342 & -3.271 & -1.604 & -0.801\\
            1.094 & 1.304 & 4.330 & 9.644 & 51.987 & -53.094 & -11.325 & -3.920\\
            0.828 & 1.124 & 1.604 & 2.203 & 4.162 & 11.675 & 54.696 & -51.980\\
	    \end{bmatrix}
	    \frac{krad}{A^2}\\
	    \Phi_0&=
	    \begin{bmatrix}
	        3.8656\\
            2.838\\
            0.798\\
            0.990\\
	    \end{bmatrix}
	    rad, \qquad
	    B=
	    \begin{bmatrix}
	        53.604 & -52.942 & -12.937 & -4.535 & -2.067 & -1.504 & 0 & -0.730\\
            3.779 & 10.918 & 52.829 & -54.752 & -9.796 & -3.963 & 0 & -1.201\\
            1.165 & 1.870 & 3.826 & 11.283 & 48.144 & -54.833 & 0 & -3.791\\
            0.706 & 0.926 & 1.338 & 2.199 & 3.731 & 11.630 & 0 & -52.863\\
	    \end{bmatrix}
	    \frac{krad}{A^2}.
	\end{align}
	Note that resistor R$_{15}$ resulted damaged during the fabrication process.

	\clearpage
	\subsection{Full optical setup}\label{supp:opticalsetup}
	
\begin{figure}[t]
        \centering
        \includegraphics[width=1\linewidth]{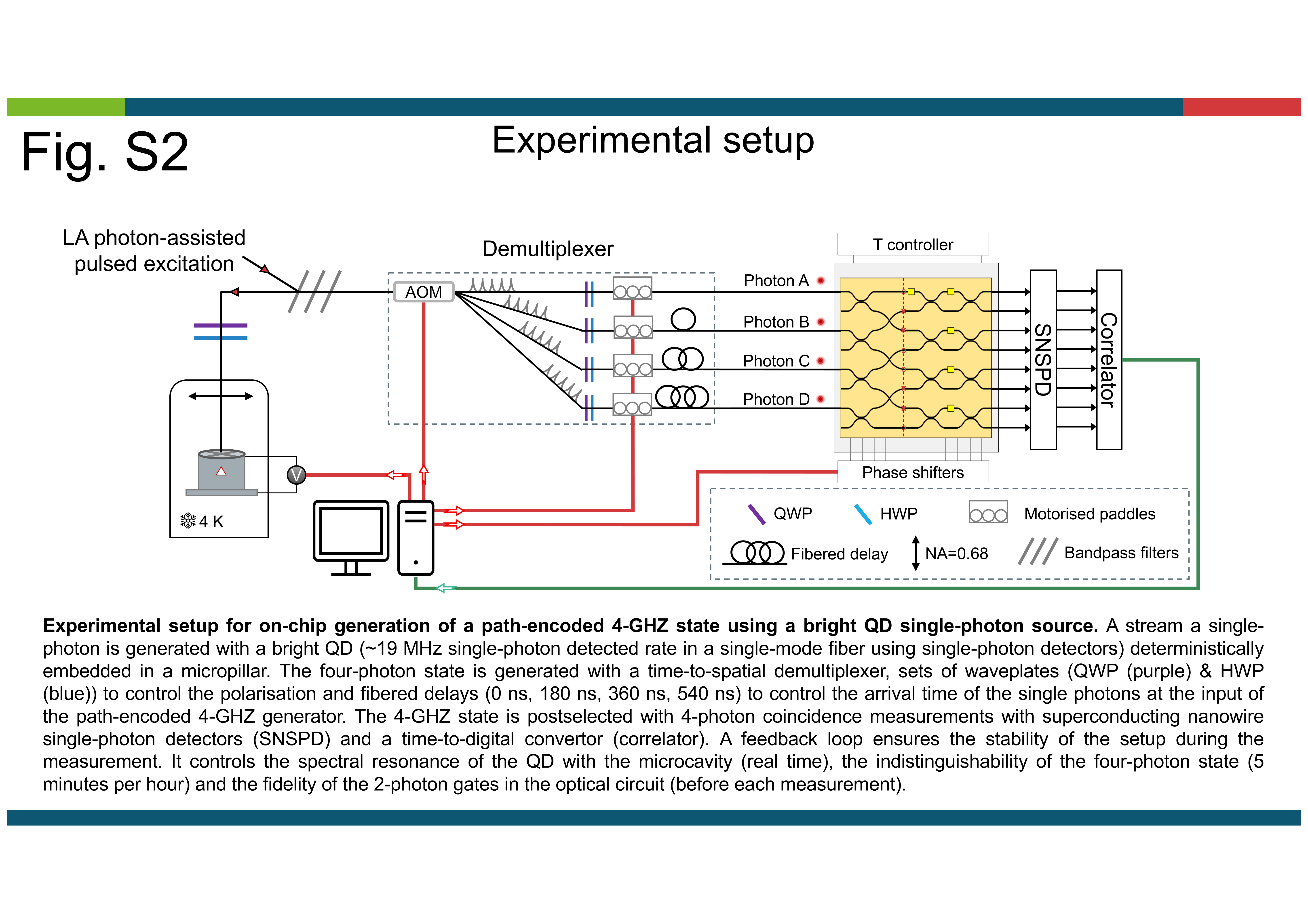}
        \caption{\textbf{Experimental setup for on-chip generation of a path-encoded 4-GHZ state using a bright QD SPS.} A stream a single-photon is generated with a bright QD ($\sim$19~MHz single-photon detected rate in a single-mode fiber using single-photon detectors) deterministically embedded in a micropillar. The four-photon state is generated with a time-to-spatial demultiplexer, sets of waveplates (QWP (purple) \& HWP (blue)) to control the polarisation and fibered delays (0 ns, 180 ns, 360 ns, 540 ns) to control the arrival time of the single photons at the input of the optical circuit. The 4-GHZ state is postselected with 4-photon coincidence measurements with superconducting nanowire single-photon detectors (SNSPD) and a time-to-digital convertor (correlator). A feedback loop ensures the stability of the setup during the measurement. It controls the spectral resonance of the QD with the microcavity (real time), the indistinguishability of the four-photon state (5 minutes per hour) and the fidelity of the projective measurements in the optical circuit (before each measurement). The temperature of the optical circuit is stabilized at $\sim295$~K with a $\sim1$~mK resolution.}
        \label{fig:experimental_setup}
    \end{figure}

	The experimental setup for on-chip generation of a path-encoded 4-GHZ state using a bright QDSPS is presented in Fig.~\ref{fig:experimental_setup}. 
	
	\subsubsection{Single-photon purity \& indistinguishability}
	\label{HOM}

	The single-photon purity of the source, defined as $\mathcal{P}_{QD}=1-g^{(2)}(0)$, characterised independently in a Hanbury Brown and Twiss setup, is $g^{(2)}(0)=0.012\pm0.001$ without spectral filtering, and can be lowered to  $g^{(2)}(0)=0.005\pm0.001$ with a narrow optical filter which does not block the single photons ($\text{FWHM}_{\text{filter}}=4\times\text{FWHM}_{\text{photon}}$), but improves the rejection of the near-resonant excitation laser. The effect of this additional spectral filter on the overall transmission of the optical setup is detailed in the loss budget (see Tab.\ref{tab:losses}). 
	
	The layout of the optical chip makes it possible to measure 4 of the 6 (AB, AC, BD and CD, see inset of Fig.~\ref{fig:Fig3}.b) two-photon wavepacket overlaps by setting the phase of all the MZIs to $\Phi_i=\pi/2$. In this configuration, we measure the 2-photon interference fringe visibility $V_{\text{HOM}}$, and compute the indistinguishability $M_s$ by taking into account the multi-photon component~\cite{ollivier2021}. Because of the time-to-spatial demultiplexing, the delay between two interfering photon is different from party to party. We present the measured indistinguishability between all accessible pairs in Tab.~\ref{tab:delay}.
	
	\begin{table}[h]
	    \centering
	    \begin{tabular}{|c|c|c|c|c|c|}
	        \hline
	        MZI Nº & Interfering photons & Delay & $V_{\text{HOM}}$ (without etalon) & $M_s$ (without etalon) \\
	        \hline
    	    MZI$_1$ & Photons A \& B & $\sim$ 180~ns & $0.91\pm0.01$ ($0.85\pm0.01$) & $0.92\pm0.01$ ($0.88\pm0.01$) \\
    	    \hline
            MZI$_2$ & Photons A \& C & $\sim$ 360~ns & $0.90\pm0.01$ ($0.82\pm0.01$) & $0.91\pm0.01$($0.85\pm0.01$) \\
            \hline
            MZI$_3$ & Photons B \& D & $\sim$ 360~ns & $0.87\pm0.01$ ($0.80\pm0.01$) & $0.88\pm0.01$($0.83\pm0.01$) \\
            \hline
            MZI$_4$ & Photons C \& D & $\sim$ 180~ns & $0.90\pm0.01$ ($0.84\pm0.01$) & $0.91\pm0.01$ ($0.87\pm0.01$) \\
            \hline
	    \end{tabular}
	    \caption{2-photon interference fringe visibility $V_{\text{HOM}}$ and indistinguishability $M_s$ of each pair of interfering photons. The multiphoton component is $g^{(2)}(0)=0.005\pm0.001$ ($g^{(2)}(0)=0.012\pm0.001$) with (without) the etalon.}
	    \label{tab:delay}
	\end{table}
	
	\subsubsection{Experimental characterisation of the loss budget}\label{Losses}
		All component optical transmissions are measured independently in order to compute the efficiencies mentioned in the main text: the collection efficiency of the optical setup $\eta_C$, the active demultiplexing efficiency $\eta_{\text{DMX}}$, the transmission of the photonic chip, mainly dominated by the insertion losses $\eta_{chip}$ and the single-photon detector efficiency $\eta_D$ are detailed in Tab.~\ref{tab:losses}. The (polarised) first-lens brightness of the source, defined as the (linearly polarised) single-photon countrate before the first optical element devided by the repetition rate of the laser, is ($\beta_{FL}\sim40\%$) $\beta_{FL}\sim50\%$. The fibered collection efficiency of the single-photon is $\eta_C$=48\% ($\eta_C$=29\%) and the overall transmission of the setup including the demultiplexing scheme and the photonic chip is $\eta$=17\% ($\eta$=10\%) without (with) the etalon filter, not including detection efficiency. The filling factor (FF) induced by the switching time of the demultiplexer is also not included in the overall efficiency, as it only reduces the effective repetition rate (RR) of the laser. The detected 4-photon coincidence rate, which must be weight by the 1/8 probability to detect one and only one photon per qubit, is thus given by:
		
		\begin{equation}
		    C_{4-photon}=RR \times FF \times (\beta_{FL} \times \eta_C \times \eta_{\text{DMX}} \times \eta_{chip} \times \eta_D)^4
		\end{equation}
	
	\begin{table}[!htb]
    \caption{Loss budget}
    \begin{subtable}{.49\linewidth}
      \centering
        \begin{tabular}{ll}
            \hline
            \multicolumn{2}{c}{\textbf{Single-photon collection efficiency} ($\eta_{C}$)}\\
    		First lens & 0.99 \\
    		Cryostat window & 0.99 \\
    		Polarisation control (QWP+HWP) & 0.96 \\
    		Coupling into SM fiber & 0.70 \\
    		Laser rejection (bandpass filters) & 0.73 \\
    		Laser rejection (etalon filter) & 0.60 \\
    		\hline 
    		\hline 
    		Degree of linear polarization & 0.70 \\
    		\hline 
        \end{tabular}
    \end{subtable}
    \begin{subtable}{.49\linewidth}
      \centering
        \begin{tabular}{ll}
            \hline
    		\multicolumn{2}{c}{\textbf{Demultiplexing efficiency}}\\
    		Optical efficiency ($\eta_{\text{DMX}}$) & 0.75\\
    		Filling factor (FF) & 0.67 \\
    		\hline
    		\hline
    		\multicolumn{2}{c}{\textbf{Photonic chip} ($\eta_{chip}$)}\\
            Insertion losses & 0.54 \\
    		\hline
    		\hline
    		\multicolumn{2}{c}{\textbf{Detector efficiency} ($\eta_{D}$)}\\
    		Mean detection efficiency & 0.65 \\
            \hline
        \end{tabular}
    \end{subtable} 
    \label{tab:losses}
    \end{table}

	\section{Modeling experimental imperfections} \label{Simulation}
	
	We build on a phenomenological model first introduced in ref.~\cite{pont2022} to identify the main sources of noise in the system. The model developed here constructs the input multi-photon state from experimentally measured metrics, and enables analysis of the calculated photon number tomography at the output of the simulated optical circuit. The simulations are implemented using the open-acess linear optical circuit development framework \textit{Perceval}~\cite{heurtel2022}. The code developed for this work and used for the numerical simulations is available at~\cite{simulation2022}.
	
	\subsection{Imperfect single-photon source} \label{imperfect sps}
	
\begin{figure}[h]
        \centering
        \includegraphics[width=0.6\linewidth]{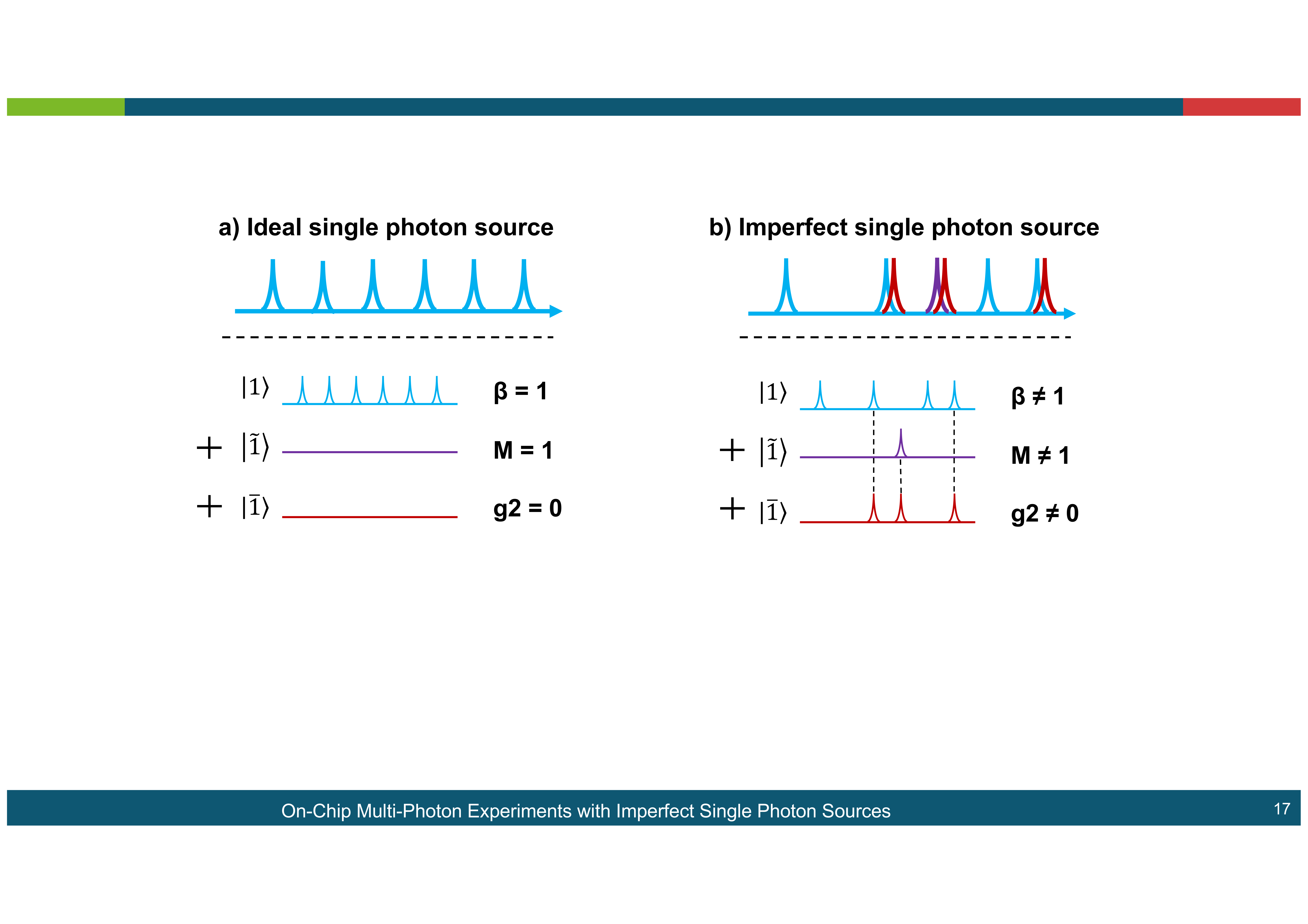}
        \caption{Representation of the state generated by an ideal single photon source (a) and an imperfect single-photon source (b). $\beta$ is the occupation probability of the QD, defined as $\beta=1-p_0$. M and g2 are the indistinguishability and purity of the single-photon state. Imperfect values are modeled respectively by  
        photons in states $\ket{\tilde{1}}$ and $\ket{\bar{1}}$.}
        \label{fig:model}
	\end{figure}

	The imperfect single-photon source is modeled by a statistical mixture of Fock states (Fig.~\ref{fig:model}). The ideal single-photon state $\ket{1}$ is mixed with distinguishable photons $\ket{\tilde{1}}$ (indistinguishability $M\ne1$), and additional noise photons $\ket{\bar{1}}$ (multiphoton component $g^{(2)}(0)\ne0$). These hypotheses together with the initial characterisation of the source ($M$ and $g^{(2)}(0)$) allow the computation of the probabilities $p_0$, $p_1$, and $p_2$ that the single-photon source emits 0, 1, and 2 photons  per time bin respectively. Here, for a bright QD based single-photon source with a LA phonon-assisted excitation scheme, we can consider that the emission process is deterministic, and set $p_0$=0. Moreover, as $g^{(2)}(0)\approx0$, we neglect all the terms with more than two photons, which sets $p_1+p_2=1$.
		
    \begin{equation}
        \label{eq:prob}
        \begin{cases}
            g^{(2)}(0)=\frac{2 p_2}{(p_1+2p_2)^2}\\
            p_1 + p_2 \approx 1\\
            p_0 \approx 0
        \end{cases}
    \end{equation}
    
    The indistinguishability of each input state \textit{i}=A,...,D is defined by the probability $x_i$ to be in a master single-photon state $\ket{1}$, shared with all other inputs, and by the probability $(1-x_i)$ to be in a subsidiary state $\ket{\tilde{1}}_i$, distinguishable from the main state, and all other subsidiary states $\ket{\tilde{1}}_j$ for $j \ne i$. The best estimators $x_i$ \textit{i}=A,...,D are computed with least-square optimization so that all the mean wavepacket overlaps between photon measured in four simulated Hong-Ou-Mandel interferometers fit the value measured experimentally. The two wavepacket overlaps $M_{BC}$ and $M_{AD}$ are not experimentally accessible, but can be computed from the $x_i$. The 4 measured mean wavepacket overlaps provide upper and lower bounds~\cite{brod2019} on $M_{BC}$ and $M_{AD}$. These bounds are included in the numerical optimization to find $x_i$ so that the result is physically acceptable. 
    
    \begin{figure}[h]
        \begin{minipage}{.45\textwidth}
        \centering
        \includegraphics[width=0.75\linewidth]{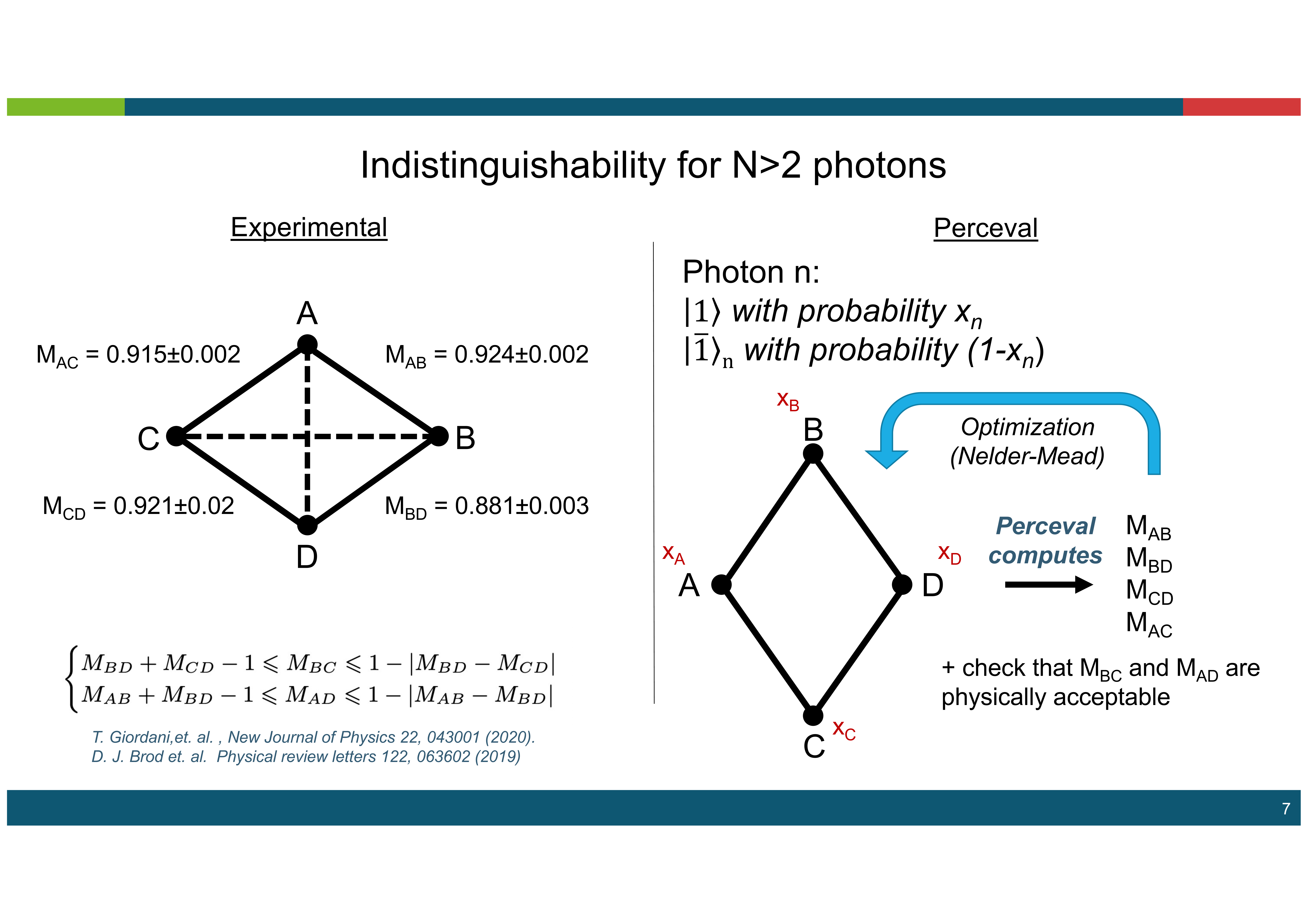}

        \end{minipage}
        \begin{minipage}{.5\textwidth}
        \label{eq:bounds}
        \begin{equation*}
        \begin{cases}
            M_{BD}+M_{CD}-1 \leqslant M_{BC} \leqslant 1 - |M_{BD}-M_{CD}|\\
            M_{AB}+M_{BD}-1 \leqslant M_{AD} \leqslant 1 - |M_{AB}-M_{BD}|
        \end{cases}
        \end{equation*}
        \end{minipage}
    \caption{Left: Graph encoding of the known experimental data. Vertices are photons, and edges are measured two-photon overlaps. Right: Bounds on the unmeasured overlaps computed from the measured 2-photon mean wavepacket overlaps.}
    \end{figure}
    
    Before demultiplexing, the input state in a given time bin generated by the imperfect single-photon source after losses can be defined by the following statistical mixture of states
    
    \begin{equation}
        \ket{\Psi}\bra{\Psi}_i=
        c_{\ket{0}}\ket{0}\bra{0}_i+
        c_{\ket{1}}\ket{1}\bra{1}_i+
        c_{\ket{\tilde{1}}}\ket{\tilde{1}}\bra{\tilde{1}}_i+
        c_{\ket{\bar{1}}}\ket{\bar{1}}\bra{\bar{1}}_i+
        c_{\ket{(1,\bar{1})}}\ket{(1,\bar{1})}\bra{(1,\bar{1})}_i+
        c_{\ket{(\tilde{1},\bar{1})}}\ket{(\tilde{1},\bar{1})}\bra{(\tilde{1},\bar{1})}_i
    \end{equation}
    
    \noindent
    where
    
    \begin{itemize}
        \item $\ket{0}$ is the vacuum.
        \item $\ket{1}_i$ is one single photon, identical for all inputs and time bins (signal). 
        \item $\ket{\bar{1}}_i$ is a one photon state, distinguishable from $\ket{1}$ and all $\ket{\bar{1}}_j$ for $j \ne i$. It models the partial distinguishability of the input state.
        \item $\ket{\tilde{1}}_i$ is a one photon state, distinguishable from $\ket{1}_i$, $\ket{\tilde{1}}_i$ and all $\ket{\bar{1}}_j$ and $\ket{\tilde{1}}_j$ for $j \ne i$. It models the unwanted multiphoton components by a two-photon state containing one single photon (signal) and one noise photon, which is distinguishable from the signal~\cite{ollivier2021}.
    \end{itemize}

    \noindent The losses in the experimental apparatus are comparable for all optical modes, so they commute with all linear optical elements, including detection efficiencies \cite{oszmaniec2018}. We can thus define an overall transmission parameter per photon, $\eta$. The probabilities for state $i$ are thus modelled as
    
    \begin{itemize}
    \begin{minipage}{0.5\linewidth}
        \item $c_{\ket{0}}^2=1-(\eta \cdot p_1 + \eta^2 \cdot p_2 + 2\eta(1-\eta) \cdot p_2)$
        \item $c_{\ket{1}}^2=\eta x_i \cdot p_1 + \eta(1-\eta) x_i \cdot p_2$
        \item $c_{\ket{\bar{1}}}^2=\eta(1-x_i) \cdot p_1 + \eta(1-\eta)(1-x_i) \cdot p_2$
    \end{minipage}
    \begin{minipage}{0.4\linewidth}
        \item $c_{\ket{\tilde{1}}}^2=\eta(1-x_i) \cdot p_1 + \eta(1-\eta)(1-x_i) \cdot p_2$
        \item $c_{\ket{(1,\tilde{1})}}^2=\eta^2x_ip_2$
        \item $c_{\ket{(\bar{1},\tilde{1})}}^2=\eta^2(1-x_i) \cdot p_2$
    \end{minipage}
    \end{itemize}
    
    The multi-photon input states generated by the demultiplexer are given by all possible combination of  the $\ket{\Psi}_i$ states for i=A,...,D, leading to $4^6=4096$ instances. Using \textit{Perceval}, the permanent of the scattering matrix corresponding to a given input state and a unitary matrix representing the optical chip (which varies depending on the configuration of the phase shifters) is computed. Each output state (number of photon per optical mode) is mapped to an outcome in the computational basis ($\ket{0}=0$ or $\ket{1}=1$ for each party). Because we use 
    threshold single-photon detectors, the probability for a detector to click is equal whatever the number of photons detected, which leads to the following mapping $(\text{number of photons per output port}) \rightarrow (\text{qubit values})  $
    \\
    \begin{align*}
        (1, 0, 1, 0, 1, 0, 1, 0) \rightarrow (0, 0, 0, 0)  \\
    (0, 1, 1, 0, 1, 0, 1, 0) \rightarrow (1, 0, 0, 0)  \\
    (0, 2, 1, 0, 1, 0, 1, 0) \rightarrow (1, 0, 0, 0) \\
    ...
    \end{align*}
    \\
    The probability of an outcome is given by the sum of the probabilities of all output states yielding that outcome, weighted by the probability of the input. The simulation is restricted to the input states with $\geq4$ photons (1041 instances), and the input states associated with a negligible probability (1e8 times smaller than the dominant term) are neglected. To run the numerical simulations presented in Tab.~\ref{tab:simu}, the parameters used, independently determined experimentally, are $g^{(2)}(0)=0.005\pm0.001$, $M_{AB}=0.924\pm0.002$, $M_{BD}=0.881\pm0.003$, $M_{CD}=0.921\pm0.002$, $M_{AC}= 0.915\pm0.002$ and $\eta=0.039\pm0.001$. The typical computation time to reconstruct the density matrix (cf. Fig.~\ref{fig:Fig2}.b) with these simulations is $\sim30$~min.

    \subsection{Imperfect preparation} \label{imperfect chip}
    
    We define the directional coupler (first row of beams-splitters, see Fig.~\ref{fig:chip}) reflectivity $\mathcal{R}$ as the fraction of optical power that, after the coupler, remains in the 
    upper waveguide.
    The target value is 0.5 (balanced coupler). In Tab.~\ref{tab:DC} we report the value of $\mathcal{R}_x$ as measured for DC$_x$ at the wavelengths of 926 nm for both H- and V-polarized lights. The reflectivity of each DC exhibits a deviation from the ideal behavior $<1\%$.
    The values reported in Tab.~\ref{tab:DC} have been measured before pigtailing output fibers. Measurements performed after this operation may be distorted by differential losses at the circuit input/output due to imperfect gluing. This effect is taken into account when computing the scattering matrix using the mean value for H- and V-polarized lights, as we do not control the polarisation entering into the chip.
    
    \begin{table}[h]
        \centering
        \begin{tabular}{|c|c|c|c|c|}
            \hline
             Polarization & DC$_1$ & DC$_2$ & DC$_3$ & DC$_4$\\
             \hline
             H & 0.499 & 0.505 & 0.490 & 0.502 \\
             \hline
             V & 0.501 & 0.505 & 0.491 & 0.504 \\
             \hline
        \end{tabular}
        \caption{Directional coupler reflectivities.}
        \label{tab:DC}
    \end{table}
    
    The imperfect preparation of the internal phase $\theta$ of the GHZ states is not taken into account in the model, as it would only affect the fidelity of the state to the target, and not the purity. The fidelity of the state to the whole class of GHZ states of the form $\ket{\text{GHZ}_4}^{(\theta)}= \left( \ket{0101}+e^{i\theta}\ket{1010} \right)/\sqrt{2}$ was calculated, and we compute a maximum fidelity of $0.863\pm0.004$ when $\theta=0.065$~rad, which falls into the error bar of the fidelity to the target. We can thus conclude that the imperfect initialisation of the internal phase is negligible.

    \subsection{Imperfect detectors \& tomography}\label{supp:projectivemeasurements}
    
	The fidelity of a projective measurement correspond to the implementation of a given projection using the tunable Mach-Zehnder interferometers (MZI). While a fine calibration of all crosstalks between the phase-shifters in the optical circuit allows to implement any gate with a near-unity fidelity, the unbalanced efficiency of the SNSPDs effectively reduces the fidelity of the projective measurement. As the SNSPD efficiency depends highly on the polarisation, it varies sporadically in the time needed to run the experiment ($\sim50$~h for the reconstruction of the density matrix) because of external noise (temperature, vibration, etc.). To counterbalance this effect, we slightly correct the calibration of each phase shifters so that the detected balance meets the target. 
	
	To characterise the fidelity of the projective measurement, we switch off 2 channels of the DMX so that only two inputs of the optical chip are used to address each MZI only on the upper (lower) input for each odd (even) parties. In this configuration the single-photon source is used as a classical source of light and the balance between the two detectors is compared to the target. The error is computed for each of the 81 measurements needed to reconstruct the density matrix. The mean error for each party before (after) re-calibration is 2.6\% (1.5\%) for party A, 5.5\% (1.4\%) for party B, 3.2\% (0.5\%) for party C, and 1.6\% (3.3\%) for party D. 
\clearpage	
	\section{Bell-like inequality measurements} \label{supp:BLI}
	
	The phase shifts associated with each Pauli projective measurement is reported in Tab.~\ref{tab:projectors}.
	
\begin{table}[h]
		\centering
		\begin{tabular}{|P{3cm}|P{4cm}|P{2cm}|P{2cm}||P{2cm}|P{2cm}|}
			\hline
			Measurement & Projector & $\chi$ & $\psi$ & $\alpha$ & $\phi$ \\[1.5ex]
			\hline
			$\sigma_x$ &  $0.707\ket{0}+0.707\ket{1}$ & $\frac{\pi}{4}$ & 0 & $0$ & $\frac{\pi}{2}$\\ [1.5ex]
			\hline
			$-\sigma_x$ & $0.707\ket{0}-0.707\ket{1}$ & $-\frac{\pi}{4}$ & 0 & $0$ & $\frac{3\pi}{2}$\\ [1.5ex]
			\hline
			$\sigma_y$ & $0.707\ket{0}+i0.707\ket{1}$  & $\frac{\pi}{4}$ & $\frac{\pi}{2}$ & $\frac{\pi}{2}$& $\frac{\pi}{2}$\\ [1.5ex]
			\hline
			$\sigma_z$ & $\ket{0}$ & 0 & 0 & $0$ & $\pi$\\ [1.5ex]
			\hline
			$-\sigma_z$ & $\ket{1}$ & $\frac{\pi}{2}$ & 0 & $0$ & $0$\\ [1.5ex]
			\hline
			$\frac{\sigma_x+\sigma_z}{\sqrt{2}}$ & $0.924\ket{0}+0.383\ket{1}$ & $\frac{\pi}{8}$ & 0 & $0$ & $\frac{3\pi}{4}$\\ [1.5ex]
			\hline
			$\frac{\sigma_x-\sigma_z}{\sqrt{2}}$ & $0.383\ket{0}+0.924\ket{1}$ & $\frac{3\pi}{8}$ & 0 & $0$& $\frac{\pi}{4}$\\ [1.5ex]
			\hline
		\end{tabular}
		
		\caption{Phase shifts to perform the projectors required by inequality in Eq.~\eqref{eq:ine}. $\chi$ and $\psi$ are the parameters characterizing the eigenstate of each operator corresponding to eigenvalue +1: $\cos\chi\ket{0}+e^{i\psi}\sin\chi\ket{1}$. $\alpha$ refers to the phase difference between the input modes of the Mach-Zehnder interferometer. $\phi$ refers to the phase difference between the arms of the Mach-Zehnder interferometer (see Fig.~\ref{fig:chip}).}
		\label{tab:projectors}
	\end{table}
	
	The measurements that are required to obtain the highest violation of Eq.~\eqref{eq:ine} in Section~\ref{sec:Bell} are the following: $M_0= \frac{\sigma_x+ \sigma_z}{\sqrt{2}}$ and $M_1= \frac{\sigma_x-\sigma_z}{\sqrt{2}}$ for party $1$, $M_0= \sigma_x$ and $M_1= \sigma_z$, for party $3$ and $M_0= -\sigma_x$ and $M_1= -\sigma_z$, for parties $2$ and $4$. The parameters $\alpha_{i}$ and $\phi_{i}$ (with $i \in (1,4)$) corresponding are shown in Tab.~\ref{tab:projectors}. As mentioned, this inequality requires $8$ expected values, which are summarized in Tab.~\ref{tab:expval}.
	
	\begin{table}[h]
		\centering
		\begin{tabular}{|P{3cm}| P{3cm}|P{3cm}|P{3cm}|P{3cm}|}
			\hline
			Operator &  Party 1 &  Party 2 & Party 3 & Party 4 \\[1.5ex]
			\hline
			$M_1^{(1)}M_1^{(2)}$ & $\frac{\sigma_x-\sigma_z}{\sqrt{2}}$  &   $-\sigma_z$ & $\mathds{1}$ & $\mathds{1}$\\ [1.5ex]
			\hline
			$M_1^{(1)}M_1^{(3)}$ & $\frac{\sigma_x-\sigma_z}{\sqrt{2}}$ & $\mathds{1}$ & $\sigma_z$ & $\mathds{1}$ \\ [1.5ex]
			\hline
			$M_1^{(1)}M_1^{(4)}$ &  $\frac{\sigma_x-\sigma_z}{\sqrt{2}}$ &  $\mathds{1}$ &  $\mathds{1}$ &   $-\sigma_z$\\ [1.5ex]
			\hline
			$M_0^{(1)}M_1^{(2)}$ & $\frac{\sigma_x+\sigma_z}{\sqrt{2}}$ & $-\sigma_z$ & $\mathds{1}$ & $\mathds{1}$ \\ [1.5ex]
			\hline
			$M_0^{(1)}M_1^{(3)}$ & $\frac{\sigma_x+\sigma_z}{\sqrt{2}}$ & $\mathds{1}$ & $\sigma_z$ & $\mathds{1}$ \\ [1.5ex]
			\hline
			$M_0^{(1)}M_1^{(4)}$ & $\frac{\sigma_x+\sigma_z}{\sqrt{2}}$ & $\mathds{1}$ & $\mathds{1}$ & $-\sigma_z$\\ [1.5ex]
			\hline
			$M_0^{(1)}M_0^{(2)}M_0^{(3)}M_0^{(4)}$ & $\frac{\sigma_x+\sigma_z}{\sqrt{2}}$ & $-\sigma_x$ & $\sigma_x$ & $-\sigma_x$ \\ [1.5ex]
			\hline
			$M_1^{(1)}M_0^{(2)}M_0^{(3)}M_0^{(4)}$ & $\frac{\sigma_x-\sigma_z}{\sqrt{2}}$ & $-\sigma_x$ & $\sigma_x$ & $-\sigma_x$ \\ [1.5ex]
			\hline
		\end{tabular}
		\caption{Measurements operators that maximize the violation of Eq.~\eqref{eq:ine}. The maximal achievable value is $6\sqrt{2} \approx 8.4853$.}
		\label{tab:expval}
		\vspace{1cm}
	\end{table}
 
    Non-classicality certification techniques which rely on the violations of given causal constraints are only valid for processes that are describable through the corresponding causal structures. For this reason, although these violations require a priori no assumption on the inner functioning of the device generating the states to certify, it is still necessary to ensure that the cause-effect relationships among the variables are the correct ones. 
    
    To draw conclusions from violating the classical bound of inequality in Eq. (\ref{eq:ine}) we need to ensure that the causal structure presented in Fig.~\ref{fig:Fig3}.a describes faithfully our experimental implementation. We do so by using part of our knowledge of the integrated circuit, depicted in Fig.~\ref{fig:Fig1} and detailed in Fig.~\ref{fig:chip}.a, thus making our conclusions \textit{semi-device independent}. To validate the causal structure we make the fair sampling assumption and we then need to ensure the two following constraints:
    \textit{(i) Measurement Independence:} there is no mutual influence among the measurement choices, and 
    \textit{(ii) Causal influence:} the parties $1$, $2$, $3$ and $4$ are independent. For condition \textit{(i)}, no observational data can exclude that measurements choices have a common cause in their past, i.e. a superdeterministic scenario can never be excluded. For condition \textit{(ii)}, our implementation exploiting an integrated photonic circuit cannot rely on space-like separation between the parties that would  enforce the condition of no direct causal influence between them by physical principles~\cite{shalm2015strong, hensen2015loophole, giustina2015significant}. 
    Hence, we use our knowledge of the apparatus, to estimate the probability that i) and ii) are satisfied by our apparatus. We identify the main a source of causal influence among the implemented measurements to be represented by thermal cross-talk. When some of the circuit resistors are heated to perform given measurements, heat dispersion affect also the other ones, introducing a correlation between the operations carried out on different photons. However, this effect can be considered negligible, as shown by the chip characterization measurements reported in the Supplementary~\ref{supp:photoniccircuit}.
    
    We thus assume that the level of causal influence of our experimental apparatus is not sufficient to reproduce the observed correlations, thus justifying that the observed violation of the Bell-like inequality certifies the presence of non-classical correlations. 

	\clearpage
	
	\section{Protocol for Quantum Secret Sharing} \label{QSS}
	We summarize in the following the 4-parties quantum secret sharing protocol described in \cite{hillery1999}, adapted to the use of our target state.
	
	\begin{enumerate}
		\item The regulator A prepares a string of 4-qubits in a GHZ state of the form: $\ket{\text{GHZ}_4}=(\ket{1010}+\ket{0101})/\sqrt{2}$, keeps qubit 1 and distributes qubits 2 to B, qubit 3 to C and qubit 4 to D.
		
		\item All four parties choose randomly a basis for measuring the state of their qubit between $\sigma_x$ and $\sigma_y$, whose eigenstates are $|\pm x\rangle$ and $|\pm y\rangle$ respectively. 
		
		\item Depending on the choice of each party, it is possible to identify the following cases:
		\begin{enumerate}
			\item All parties measure the qubits according to the same basis, e.g. $\sigma_x$. Accordingly, the state of the qubits can be written as an equal superposition of all terms that contain only an \underline{\textit{even}} number of $\ket{-x}$ states.
			Thanks to this, B, C and D can determine the result of A by comparing their results: if they measured an even number of $\ket{-x}$ states then A would find its qubit in the state $\ket{x}$, and vice versa.
			
			\item One of the parties measures on a different basis with respect to all others. Accordingly, the 4-qubits state can be written as an equal superposition of \underline{\textit{all}} 16 possible combinations of the measurements eigenstates and, due to this, the joint knowledge of the results of B, C and D produces no information on A's.	
			
			\item Two parties sharing correlated qubits, e.g. A and C, measure on $\sigma_x$, while the others measure on $\sigma_y$. In this case the overall state can be written as a superposition of all terms that contain only an \underline{\textit{odd}} number of negative states ($\ket{-x}$ or $\ket{-y}$). Similarly as in case (a), parties B, C and D can determine A's result by counting the number of negative states they found.
			
			\item Two parties sharing anti-correlated qubits, e.g. A and B, measure on $\sigma_x$, while the others measure on $\sigma_y$. In this case the overall state can be written as a superposition of all terms that contain only an \underline{\textit{even}} number of negative states. Again, the result of A can be retrieved by those of B, C and D.
			
		\end{enumerate}
		
		\item B, C and D announce publicly their measurement basis. A compares their choices with its own, and will communicate to B, C and D to discard the results belonging to case (b), while will make public its basis choice for the other cases. Among all possible 16 combinations of choices, 2 belong to case (a), 8 belong to case (b), 2 belong to case (c) and 4 belong to case (d). So, on average, there is 50\% probability to obtain a useful result.
		
		\item The qubit sharing and measurement process is iterated 2N times, which leads to the establishment of a sifted key of length N.
		
		\item A part of the sifted key can be exchanged publicly and used for quantum bit error rate (QBER) evaluation. Similarly as in BB84 QKD scheme, communication eavesdropping can be detected by measuring anomalously high values of QBER.
		
		\item If the QBER is within the agreed threshold, the remaining part of the string can be used for error correction and privacy amplification.
	\end{enumerate}
	
\end{document}